\documentclass[%
 aip,
 amsmath,amssymb,
 reprint,%
]{revtex4-1}

\usepackage{graphicx}
\usepackage{dcolumn}
\usepackage{bm}
\usepackage{natbib}
\usepackage{subfig}
\usepackage{amssymb}
\usepackage{amsmath}
\usepackage{gensymb}
\usepackage{setspace}
\usepackage{xcolor}
\usepackage[hidelinks]{hyperref}
\usepackage[all]{hypcap}
\usepackage[capitalise]{cleveref}
\renewcommand{\figurename}{FIG.}

\draft
\begin{document}
\preprint{AIP/123-QED}
\title{Spectral analysis of flow and scalar primitive variables in near and far laminar wake of an elliptic cylinder}

\author{V. Pulletikurthi}
 \affiliation{Department of Mechanical Engineering, Indian Institute of Technology Madras, India}
\author{I. Paul}%
 \email{immanuvelpaul@gmail.com}
\affiliation{Department of Applied Mechanics, Indian Institute of Technology Madras, India}

\author{K. A. Prakash}
\affiliation{Department of Applied Mechanics, Indian Institute of Technology Madras, India}%

\author{B.V.S.S.S Prasad}
\affiliation{Department of Mechanical Engineering, Indian Institute of Technology Madras, India}%

\date{\today}

\begin{abstract}
We analyze the primitive variables of fluid flow and scalar fields through fast Fourier transform (FFT) in the near and far wake of an elliptic cylinder. Numerical simulation of flow and scalar fields behind an elliptic cylinder of axis ratio  0.4 at a Reynolds number of 130 is performed. The semi-major axis of the elliptic cylinder is kept perpendicular to the incoming flow, where the fluid flow is two-dimensional and the Prandtl number is 0.71. The scalar is injected into the flow field by means of heating the cylinder continuously. The simulation is run for a long time to show that the secondary vortex street is a time-dependent phenomenon. Three distinguishable flow and scalar regions are observed in the wake of the cylinder. This study reveals the presence of low-frequency structures besides the primary shedding structures in linear, transition and saturation regions of temporal wake development. We show that the spectral source of the primary frequency is the saturated state of the temporal wake development, while its physical source is the periodic arrangement of structures of primitive variables, which inhibits the transmutation of their wavelength. On the other hand, the secondary low frequency is embedded in the transitional developing stage of the wake and its physical source is the chaotic behaviour of the transition process, which aids in the transmutation of the wavelength of the structures. Our spectral analysis also reveals that the scalar is predominately carried by the streamwise velocity and the pressure throughout the wake.
\end{abstract}

\maketitle

\section{Introduction}
Understanding the evolution of flow structures behind a bluff body is important for gaining the fundamental knowledge of fluid flow as well as for applications such as drag reduction and heat transfer enhancement. One of the important observations concerning flow structures behind a bluff body is the low-frequency unsteadiness in the far-wake of a cylinder. 

\citet{roshko1954development} was the first to observe the low-frequency unsteadiness in the very far wake of a circular cylinder for $x/D> 150 $ (here, $x$ is the streamwise distance and $D$ is the diameter of the cylinder). Later, \citet{tritton1959experiments} and \citet{berger1964determination} observed that the two different modes of vortex shedding relate to the flow structures of high and low frequencies for $Re < 180$. These studies concluded that the flow non-uniformities induce low-frequency unsteadiness.

Unlike the circular cylinder, the wakes of flat plates have the presence of low-frequency unsteadiness even in the near wake. The study of \citet{najjar1998low} demonstrated that the interaction of von-Karman vortex street in the wake of a flat plate is the reason behind the low-frequency unsteadiness. However, it should be mentioned here that their observation is based on the drag coefficient ($C_D$) curve, and therefore their mechanism of low-frequency unsteadiness is based on the $C_D$. For instance, they stated that the constant back and forth switching between the shedding cycles of low and high mean drag causes low-frequency unsteadiness. Note that their explanation could not be extended to 2D wakes of elliptic cylinders as the drag over both geometries are different.

Recent research carried out on elliptic cylinder wakes revealed the presence of low-frequency unsteadiness in the spectra of cross-stream velocity component even in the near wake. \citet{john04} were the first to extensively analyze the wakes of elliptic cylinders for low-frequency unsteadiness. They carried out two-dimensional simulations for various axis ratio (which is defined as the ratio between the semi-major and minor axes) and Reynolds numbers. They concluded that the low-frequency unsteadiness is due to a two-dimensional instability. The other works on the elliptic cylinder (\citet{radi2013experimental} and \citet{Thom14}) also echo the conclusion of \citet{john04}.

Very recently, \citet{pauls16} attempted to propose a new explanation for the low-frequency unsteadiness in the two-dimensional wakes of elliptic cylinders. They proposed that the secondary frequencies are due to temporal wake development. Therefore, they split the velocity signal into different segments and showed that the secondary low frequency is from the wake transition. However, their way of splitting the signal is not accurate and the root cause (physical source) of low-frequency unsteadiness remains unknown. In this paper, we attempt to fill these gaps by performing numerical simulation of flow past an elliptic cylinder at $Re=130$ for which the flow is reported to be two-dimensional by \citet{Thom14}. In addition, we also consider the scalar field and we look for the low-frequency unsteadiness in the temperature spectra. The overall objective of this paper is to establish the physical and spectral sources of low-frequency unsteadiness in the primitive variables of fluid flow and scalar within the laminar regime.

This paper is structured as follows. Section \ref{governeq} provides the numerical details of the simulation. Subsequently, the results are discussed in sections \ref{temporal}, \ref{fast}, \ref{spectral}, \ref{physical} and conclusions are reported in section \ref{conc}. We first discuss the temporal evolution of the fluid flow and scalar fields using the vorticity and temperature contours in section \ref{temporal}. We then analyze the fluid flow and scalar primitive variables using fast Fourier transform (FFT) to observe low-frequency unsteadiness in section \ref{fast}. Following this section, we provide the spectral and physical source of the low-frequency unsteadiness. The spectral source is revealed through an improved signal decomposition method based on the complex demodulation technique in section \ref{spectral}. The physical source is explained through instantaneous contours of velocity and temperature in section \ref{physical}. Finally, we summarize the results in section \ref{conc}. 

\section{\label{governeq}Governing Equations and Numerical Methodology}

\indent We simulate the fluid flow and scalar fields using one of the Cartesian grid methods called the Immersed Boundary Method (IBM). We followed the \citet{su07} algorithm with two adaptations. Firstly, we have used the method by \citet{Brown01} for a second-order convergence in pressure by integrating the Navier-Stokes equations using a second-order projection scheme. The second alteration is to provide a wider stability region to Dirac delta function by integrating Eulerian and Lagrangian variables through a 4-point regularized Dirac delta function described in \citet{Shin08}. In this method, the solid body is immersed in the flow through a forcing function. In detail, the flow domain (Eulerian domain) is developed first and solid body is immersed on the flow domain using Lagrangian marker points. Dirac delta function integrates the Eulerian domain with the Lagrangian marker points (i.e. solid body in flow domain). The forcing term appears in the momentum as well as energy equations. Thus the governing equations of the numerical simulation are :
\begin{equation}
\nabla\cdot \textbf{u} = 0 
\label{1}
\end{equation}
\begin{equation}
\frac{\partial \textbf{u}}{\partial \tau} + (\textbf{u}\cdot\nabla) \textbf{u} = -\nabla p + \frac{1}{\mathit{Re}}{\nabla}^2 \textbf{u} + \mathbf{f}
\label{2}
\end{equation}
\begin{equation}
\frac{\partial \textbf{T}^\textbf{{*}}}{\partial \tau} + (\textbf{u}\cdot\nabla) \textbf{T}^\textbf{{*}} = \frac{1}{\mathit{Re}\mathit{Pr}}{\nabla}^2 \textbf{T}^\textbf{{*}} + \mathbf{q}
\label{3}
\end{equation}
where \textbf{u} is the non-dimensional velocity vector (\textbf{u} = $\frac{u}{u_{\infty}}$, u - dimensional velocity vector in $ms^{-1}$, $u_{\infty}$ - free stream velocity in  $ms^{-1}$), p is the non-dimensional pressure ($p = \frac{p}{\rho u^{2}_{\infty}}$, $p$ - dimensional pressure in $\frac{N}{m^{2}}$, $\rho$ - dimensional density in $\frac{kg}{m^{3}}$), and $\tau$ is non-dimensional time ($\tau = \frac {tu_{\infty}}{D_{H}}$, $t$ - dimensional time in $s$ and $D_{H}$ is hydraulic diameter which is defined as  $D_{H} = \frac{4A}{P}$, A is the area and P is the perimeter of the elliptic cylinder). The Reynolds number ($Re$) is defined as $Re = \frac{\rho u_\infty D_{H}}{\mu}$, where $\mu$ is viscosity of fluid. Hydraulic diameter ($D_{H}$) is taken as the characteristic length. Prandtl number, $Pr = \frac {\mu c_{p}}{k}$, where $k$ is coefficient of thermal conductivity, $c_{p}$ is specific heat capacity. In Equations \eqref{2} and \eqref{3}, the terms $\mathbf{f}$ and $\mathbf{q}$ are called the forcing terms for the momentum and energy equations respectively. These terms are associated with the IBM, and are used to represent the solid boundary in the Eulerian domain. The second-order central differencing scheme is employed to discretise the diffusive and convective terms in the governing equations. The Euler integration method is used to discretise the temporal term. The spatial dimensions of computational domain are given as $-8D_{H}\leq x \leq 100D_{H}$ and $-8D_{H}\leq y \leq 8D_{H}$.
\begin{figure}
\centering
\subfloat{\includegraphics[trim = 0mm 8cm 0.3mm 8cm,clip,width=0.5\textwidth]{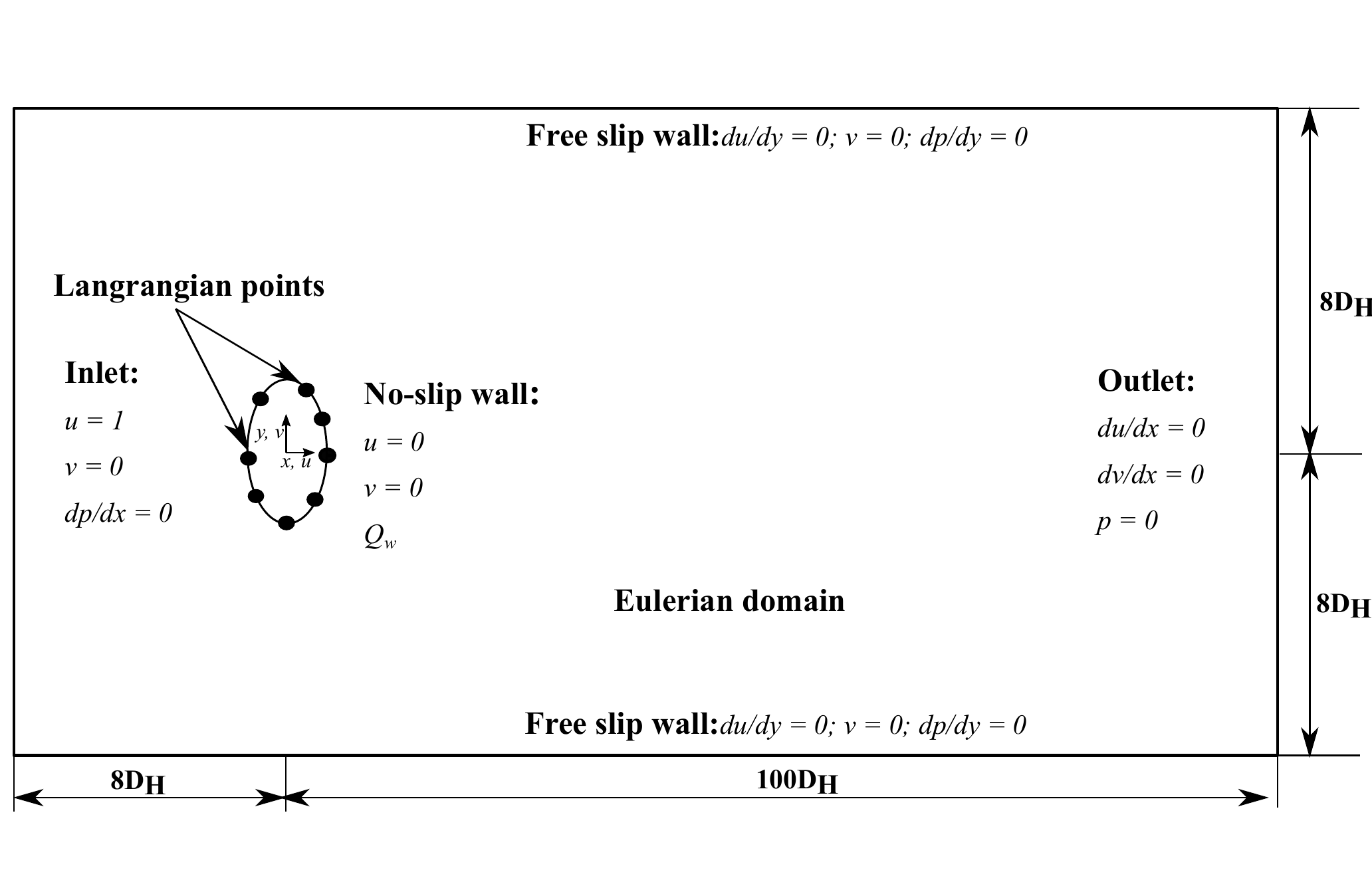}}
\caption{Computational domain}
\label{comp_domain}
\end{figure}

The computational domain is shown in Fig. \ref{comp_domain}. Uniform streamwise velocity profile is prescribed at the inlet and fully developed velocity and zero gauge pressure conditions are given at the outlet of the domain. Free slip wall boundary conditions are maintained for the top and bottom of the domain. No slip and constant heat flux ($Q_{w}$) boundary conditions are imposed for velocity and temperature on the wall of the elliptic cylinder. A Lagrangian domain is used to represent the cylinder using 315 marker points. A non-uniform grid with 766 and 217 points is used for the Eulerian domain along the $x$ and $y$ directions respectively, which makes approximately 166222 grid points. A uniformly spaced grid is embedded around the elliptic cylinder for the effective use of forcing function and to resolve the shear layer effectively. The size of the uniform grid sized computational domain is $-1D_{H}\leq x \leq1D_{H}, -1D_{H}\leq y \leq 1D_{H}$ with the mesh size of $\Delta x$ = $\Delta y = 0.01$.
Further details about the implementation of code, validation, grid and domain independent studies can be found in the previous works of the authors (\citep{paul13,Paul14a,Paul14b,paul2013effect}).

\section{Temporal evolution of flow and scalar field}\label{temporal}

\begin{figure*}[htpb]
\centering
\mbox{\subfloat[\label{Inst_vort_a}]{\includegraphics[trim =  0.1cm 5cm 2cm 6cm,clip,width=0.34\textwidth]{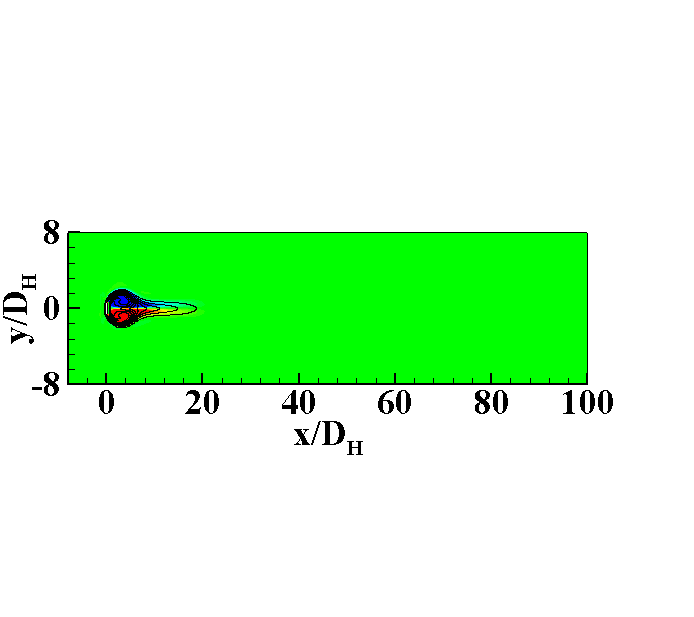}}\
\subfloat[\label{Inst_vort_b}]{\includegraphics[trim =  0.1cm 5cm 2cm 6cm,clip,width=0.34\textwidth]{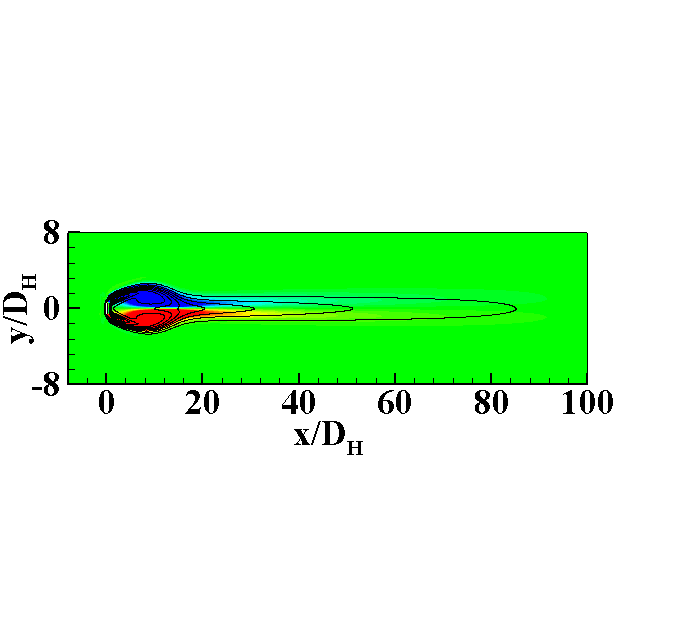}}\
\subfloat[\label{Inst_vort_c}]{\includegraphics[trim =  0.1cm 5cm 2cm 6cm,clip,width=0.34\textwidth]{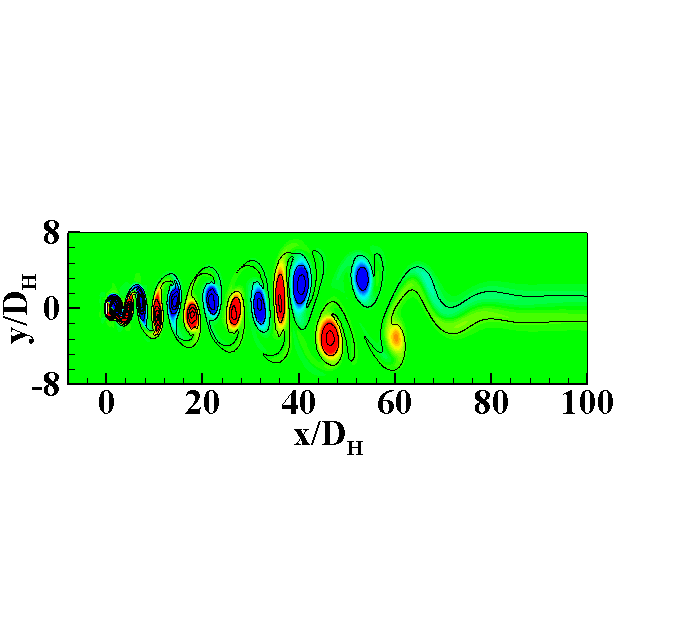}}}\
\mbox{\subfloat[\label{Inst_vort_d}]{\includegraphics[trim =  0.1cm 3cm 2cm 7.5cm,clip,width=0.34\textwidth]{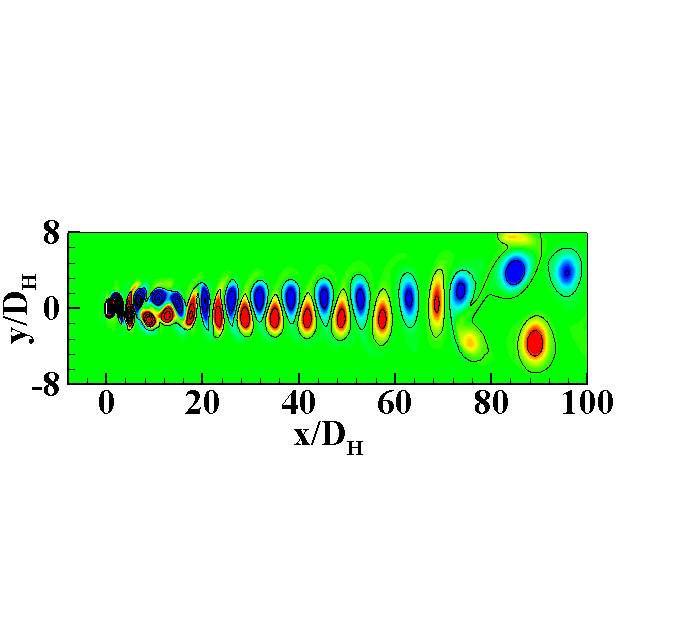}}\
\subfloat[\label{Inst_vort_e}]{\includegraphics[trim =  0.1cm 2cm 2cm 7.5cm,clip,width=0.34\textwidth]{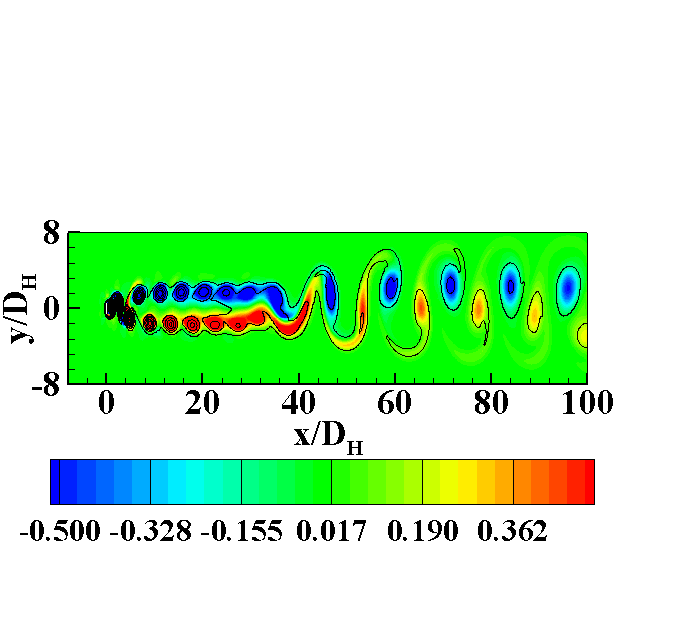}}\
\subfloat[\label{Inst_vort_f}]{\includegraphics[trim =  0.1cm 4cm 2cm 9cm,clip,width=0.34\textwidth]{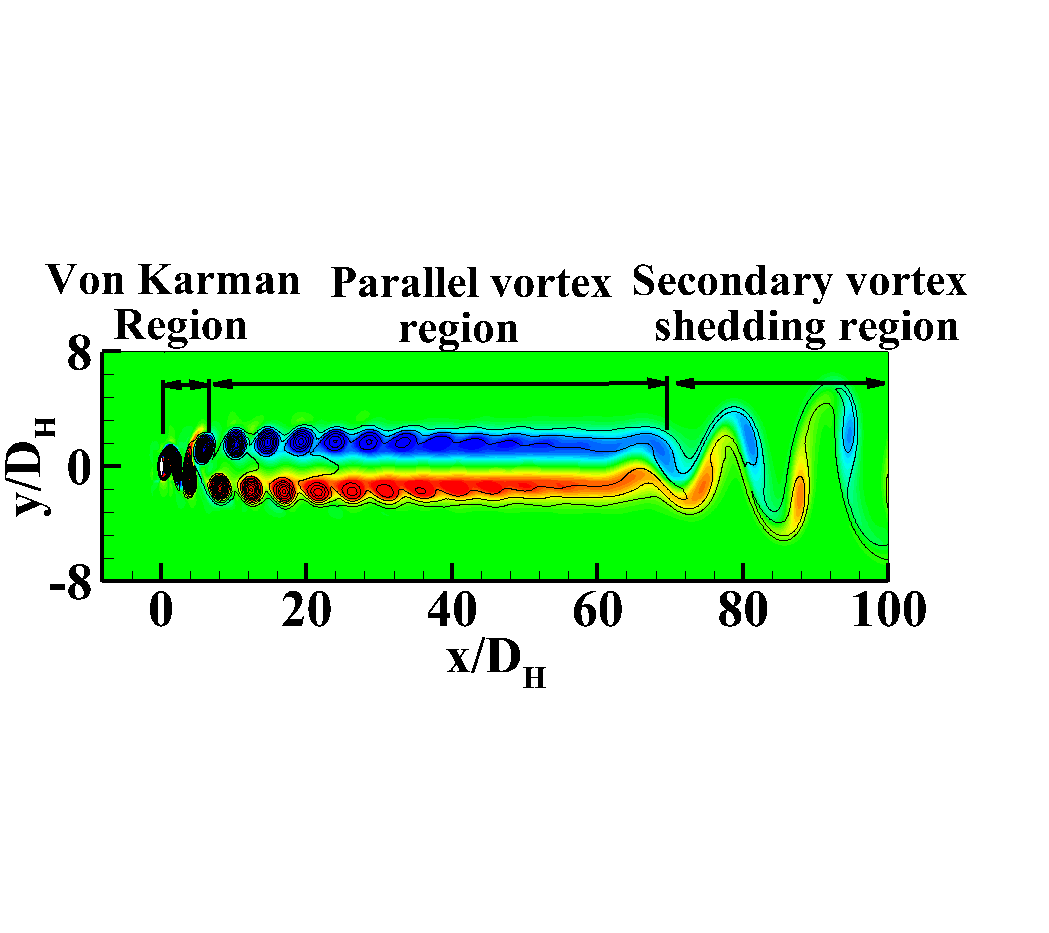}}}
\caption{Instantaneous vorticity contours(flood) and temperature(contour line level - 0.005) at various non dimensional time instincts a) $\tau$ = 25, b) $\tau$ = 105, c) $\tau$ = 180, d) $\tau$ = 223, e) $\tau$ = 460 and f) $\tau$ = 740}
\label{Inst_vort}
\end{figure*}

We first visualize the flow and scalar field under consideration through vorticity and temperature contours. Figure \ref{Inst_vort} shows the temporal evolution of vorticity and temperature contours behind the elliptic cylinder. The first thing to note from this figure is that the wake state is unsteady and it undergoes a series of transition processes. At the initial stage of the simulation, two standing vortex bubbles behind the cylinder can be noted in Figs. \ref{Inst_vort_a} and \ref{Inst_vort_b}. These standing bubbles are highly unstable and they start shedding after a certain time of the simulation as shown in Fig. \ref{Inst_vort_c}. When this shedding happens, the resulting vortex pattern downstream of the cylinder appears to be that of the von-Karman vortex street with a periodic arrangement of positive and negative vortices. 

Various studies (\citep{Durgin71,Tsuboi85,Funakoshi94}) have shown that the von-Karman vortex street is unstable for $h/a$ (where $h$ is the vertical distance between the vortices of opposite signs, and $a$ is the horizontal distance between the vortices of similar sign) ratio greater than 0.365. The measured values of $h/a$ for Fig. \ref{Inst_vort_c} exceed the value of 0.365 around $x/D_{H} \approx $10. Hence, the von-Karman vortex street of Fig. \ref{Inst_vort_c} is unstable. Consequently, the primary vortex street (i.e. the von-Karman vortex street) breaks down around $x/D_{H}\approx$ 8 and transforms into two parallel rows of vortices of same sign in the region of $8\leq x/D_{H} \leq 15$ in Fig. \ref{Inst_vort_d}. The process by which the primary vortex street loses its periodic nature is called the anisotropic extension (\citet{Funakoshi89}). \citet{pauls16} have shown that this anisotropic extension process happens due to the domination of shear which stretches the individual vortices along the direction of flow, thus forming rows of similar sign vortices.

The rows of parallel vortices are also unstable in space and they tend to merge with each other and shed periodically. The vortex street that results from the rows of parallel similar sign vortices is called the secondary vortex street which can be seen after $x/D_{H}>20$ in Fig. \ref{Inst_vort_d}. The region of parallel vortices in Fig. \ref{Inst_vort_d} continues to grow in size along the streamwise direction as the simulation time progresses further as seen in Fig. \ref{Inst_vort_e}. Eventually, the wake state reaches a dynamic steady state meaning that the wake no longer undergoes a further temporal transition in Fig. \ref{Inst_vort_f}.

Three distinct regions downstream of the cylinder can be observed in Fig. \ref{Inst_vort_f} of the dynamic steady state wake. The terminologies of these regions are in-line with the study of \citet{pauls16}. First, there exists a von-Karman region in the immediate downstream of the cylinder for $0\leq x/D_{H} \leq 5$ where the primary von-Karman vortex street exists with its periodic arrangement of vortices. Following this, there exists a parallel vortex region for $5 \leq x/D_{H} \leq 70$ which is occupied by rows of similar sign vortices aligning side by side. Finally, for $x/D_{H} >$70, we have a secondary shedding region where the vortices once again exhibit a periodic arrangement. The temporal wake development, as depicted in Fig. \ref{Inst_vort}, also reveals that it involves structures (or vortices or motions) of diverse sizes.

The temporal evolution of the scalar field is shown in the same Fig. \ref{Inst_vort} as isothermal lines on top of the flooded contours of vorticity. The scalar field is similar to that of the vorticity as the temperature is a passive scalar in this study. The scalar is trapped in the vortices and they follow the path of the carrier flow. Therefore, the thermal wake also contains the aforementioned three regions in the downstream of the heated elliptic cylinder.

\section{Fast Fourier Transform of the primitive variables of flow and scalar fields}\label{fast}
Having understood the temporal evolution of the momentum and thermal wakes, we move on to analyse the primitive variables of the wakes in this section. Since we are considering two-dimensional flow and scalar fields, their primitive variables are: (i) streamwise velocity, $u$, (ii) cross-streamwise velocity, $v$, (iii) pressure, $p$, and (iv) temperature, $T$. In this section, we perform fast Fourier transform (FFT) of these variables for their primary and secondary frequencies. To this end, we measure signals of these primitive variables at 50 equally spaced locations along the centerline of the wake through virtual probes.
\subsection{Cross-streamwise velocity}

\begin{figure*}[htpb]
\centering
\mbox{\subfloat[\label{vu_fft4}] {\includegraphics[trim =  0.4cm 0.4cm 0.2cm 0.1cm,clip,width=0.4\textwidth]{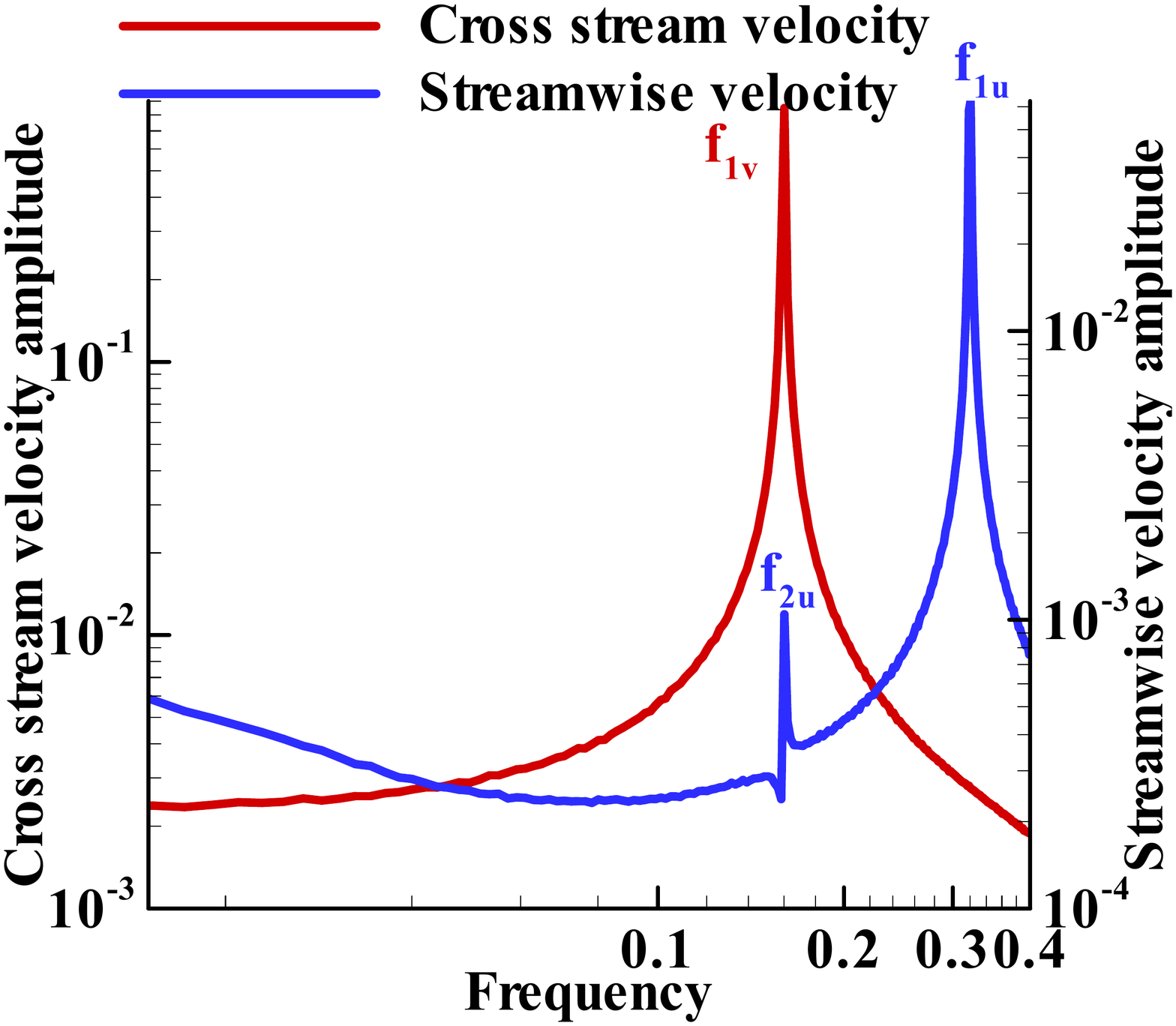}}\quad \subfloat[\label{vu_fft22}]{\includegraphics[trim =  0.4cm 0.35cm 0.2cm 0.1cm,clip,width=0.4\textwidth]{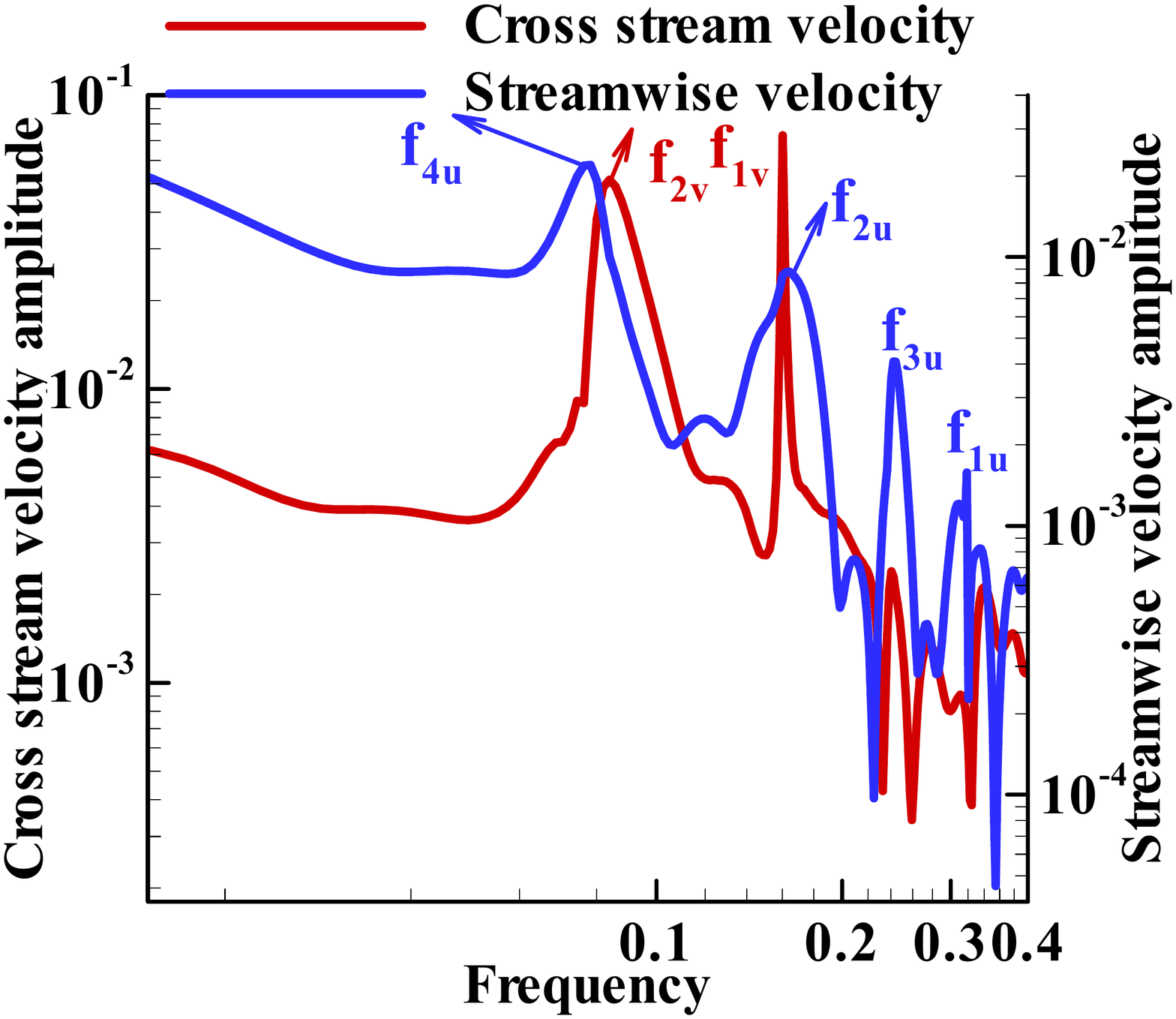}}}
\mbox{\subfloat[\label{vu_fft62}] {\includegraphics[trim = 0.4cm 0.4cm 0.2cm 0.1cm,clip,width=0.4\textwidth]{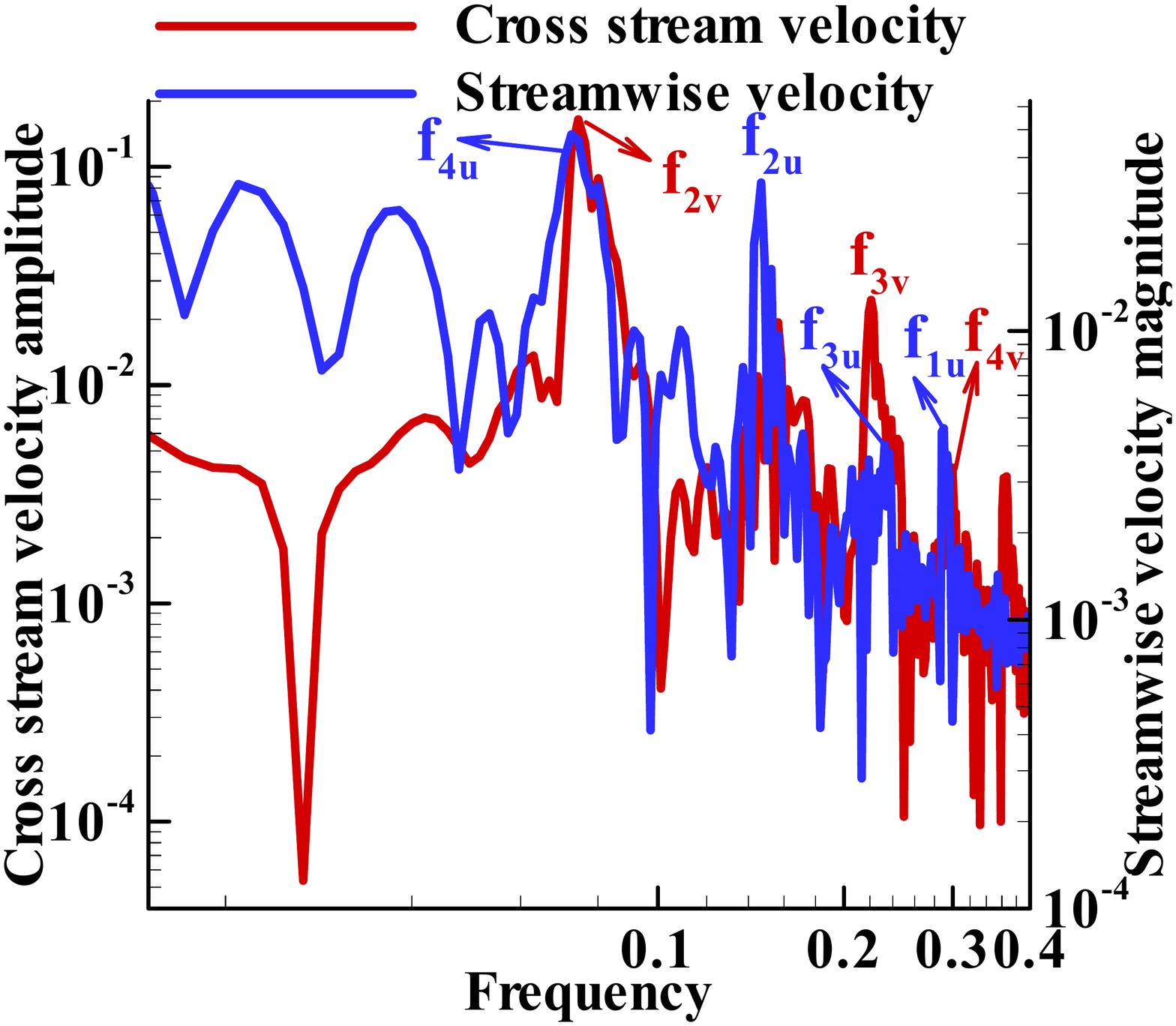}}\quad \subfloat[\label{graph_vu}] {\includegraphics[trim = 1.5cm 1.2cm 1.5cm 1.5cm,clip,width=0.4\textwidth]{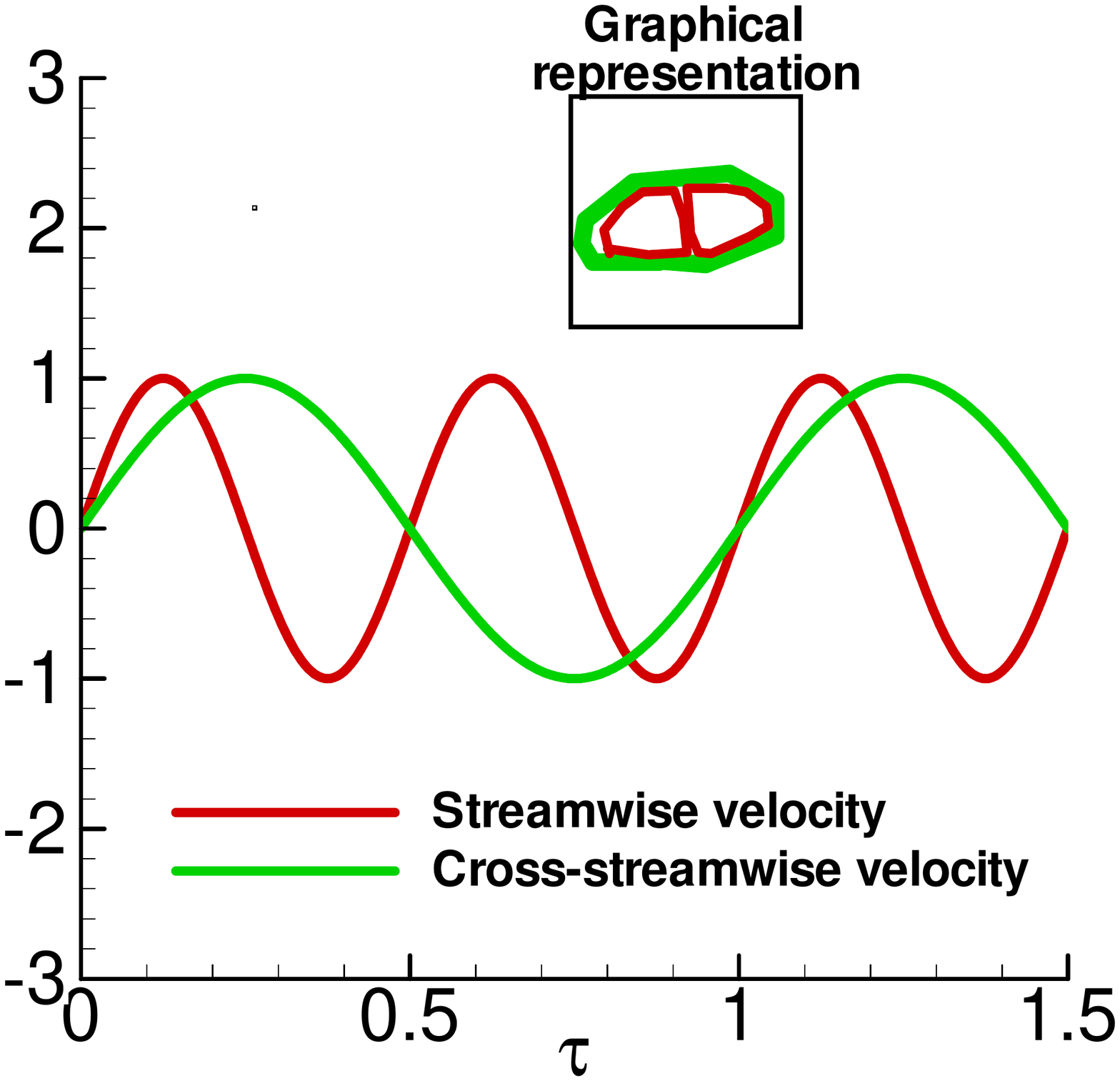}}}
\caption{FFT of cross stream velocity and streamwise velocity at: a) ($x/D_{H}$,$y/D_{H}$) = (4,0), b) ($x/D_{H}$,$y/D_{H}$) = (22,0)  and c) ($x/D_{H}$,$y/D_{H}$) = (62,0), d) Graphical representation of the streamwise and cross-streamwise velocity signals.}
\label{vu_fft}
\end{figure*}

The literature on elliptic cylinder \citet{john04} and \citet{pauls16} have used predominately the cross-stream velocity as their primitive variable. Therefore, we first look at the FFT of $v$-velocity. We present the FFT of $v$-velocity for the von-Karman, parallel vortex and secondary shedding regions in Figs. \ref{vu_fft4}, \ref{vu_fft22} and \ref{vu_fft62} respectively in red-coloured lines.

In the von-Karman region, the FFT of $v$-velocity exhibits a clearly-defined single frequency which we call it as the primary $v$-frequency ($f_{1v}$) whose value is 0.1598 according to Fig. \ref{vu_fft4}. This observation suggests that the $v$-velocity in the immediate downstream of the cylinder is insensitive to the temporal wake development as different sized structures noted in Fig. \ref{Inst_vort} are not reflected in \ref{vu_fft4}. However, the cross-streamwise velocity does become sensitive to the temporal wake development by the starting of the parallel vortex region with an appearance of a secondary low $v$-frequency ($f_{2v}$) in figure \ref{vu_fft22} along with $f_{1v}$ although the energy content of $f_{2v}$ is low. Here, the flow structures with frequency $f_{1v}$ are generated in the von-Karman region and get convected into the parallel vortex region, while the temporal wake development acts as a source for $f_{2v}$. The exact way as to how the wake development induces $f_{2v}$ is studied in sections \ref{spectral} and \ref{physical}.

The secondary $v$-frequency (i.e. $f_{2v}$) grows in its energy content along the parallel vortex region and obtains an energy content equal to that of the primary $v$-frequency (i.e. $f_{1v}$) around $x/D_{H}\approx$22 as shown in Fig. \ref{vu_fft22}. Further downstream of this point within the parallel vortex region, the energy content of $f_{2v}$ continues to grow while the opposite happens for $f_{1v}$. By the end of the parallel vortex region, the primary $v$-frequency completely vanishes and frequencies higher than $f_{1v}$ start to appear along with the secondary low $v$-frequency ($f_{2v}$).

Although the frequencies such as $f_{3v}$, $f_{4v}$ and $f_{5v}$ are present in the secondary shedding region (see Fig. \ref{vu_fft62}), they are not the dominant frequencies and their appearance keeps changing with respect to the streamwise distance within the secondary shedding region. From this, we can state that the secondary low $v$-frequency is the representative frequency of the secondary vortex street although its origin is the parallel vortex region. Also, the value of the secondary $v$-frequency (i.e. 0.07629) is not a linear multiple of the primary $v$-frequency. This evidence suggests that the secondary vortex street is not due to vortex pairing (\citet{john04} and \citet{pauls16}). These observations are in-line with the similar results presented in the studies of \citet{pauls16} and \citet{john04} although they did not associate the primary and secondary $v$-frequencies with respect to different regions of the flow field.

\subsection{Streamwise velocity}
As mentioned earlier, the previous studies on the elliptic cylinder wakes have used only the $v$-velocity for computing FFT. No information is available on the FFT of other primitive variables. This motivates our next analysis concerning the second primitive variable (i.e. streamwise velocity, $u$) whose FFT is also plotted in Figs. \ref{vu_fft4} - \ref{vu_fft62} using blue-coloured lines. 

Surprisingly, the FFT of $u$-velocity contains a higher primary $u$-frequency ($f_{1u}$) and a lower secondary $u$-frequency ($f_{2u}$) even in the von-Karman region as shown in Fig. \ref{vu_fft4}. This result is indeed interesting as the von-Karman region is generally believed to be unaffected by the temporal wake development as seen in the previous subsection. However, the FFT analysis of the streamwise velocity shows that the von-Karman region also seems to be sensitive to the temporal wake development and the way it causes secondary low frequency is explained in sections \ref{spectral} and \ref{physical}. Here, we also note that the value of $f_{1u}$ ($\approx 0.32$) is double to that of $f_{2u}$. We surmise that each of $v$-velocity structure of the von-Karman region contains two structures of $u$-velocity as shown in the inset figure of Fig. \ref{graph_vu} where the green colour structure is that of $v$ and the red one is that of $u$. As a result, when a $v$-velocity structure completes one cycle, the $u$-velocity completes two cycles, thus the frequency of $u$ is twice to that of $v$. This explanation is schematically given in Fig. \ref{graph_vu}. One might argue if such a configuration of two $u$ structures in one $v$-structure exists in the actual flow field or not. This question is answered when we present the physical source of frequencies in section \ref{physical}. Yet, it is quite interesting to note that a secondary low $u$-frequency exists even in the von-Karman region.

The two $u$-frequencies (i.e. $f_{1u}$ and $f_{2u}$) of the von-Karman region (which are not due to wake development) get convected into and survive in the parallel vortex region similar to the survival of $f_{1v}$ in the $v$-velocity. Besides these two convected $u$-frequencies from the upstream, the temporal wake development also gives rise to two additional $u$-frequencies (i.e $f_{3u}$ and $f_{4u}$) in the parallel vortex region. As a result, the FFT of $u$-velocity at $x/D \approx$22 has four frequencies as seen in Fig. \ref{vu_fft22}.

Unlike the $v$-velocity secondary shedding region where the frequency from the von-Karman region $f_{1v}$ vanished, it survives here for the $u$-velocity along with the presence of $f_{2u}$ which is generated in the von-Karman region (see Fig. \ref{vu_fft62}). Besides $f_{2u}$, the frequencies which are convected from the parallel vortex region (i.e. $f_{3u}$ and $f_{4u}$) also survive. As a result, the four $u$-frequencies of the parallel vortex region get transformed into the secondary shedding region although their identification becomes cumbersome in the secondary shedding region.
\subsection{Velocity magnitude}
\begin{figure*}[htpb]
\centering
\mbox{\subfloat[\label{p_vmag4}] {\includegraphics[trim = 2cm 0.4cm 2cm 1cm,clip,width=0.34\textwidth]{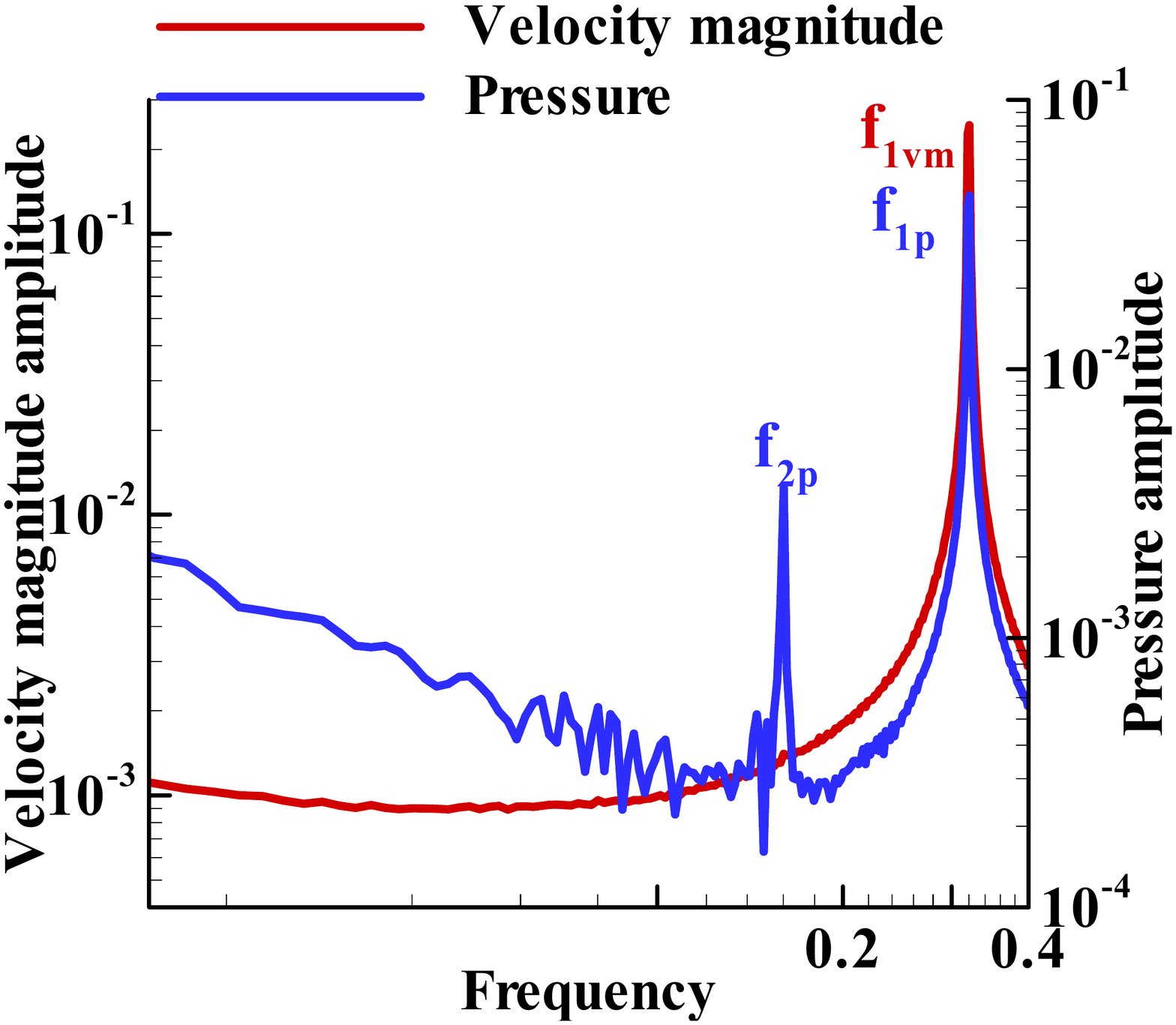}}\subfloat[\label{p_vmag22}] {\includegraphics[trim = 2cm 0.4cm 2cm 1cm,clip,width=0.34\textwidth]{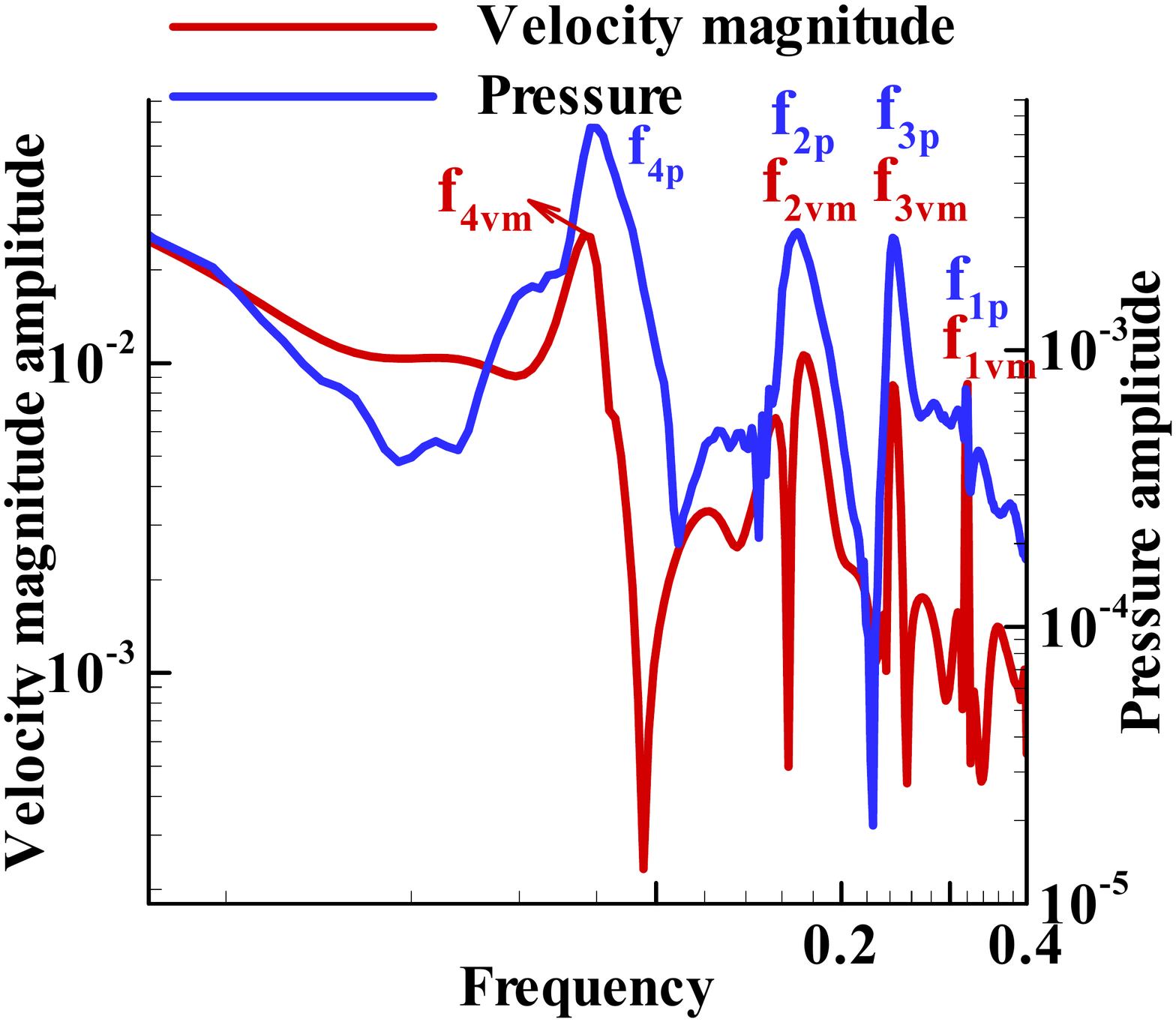}}
\subfloat[\label{p_vmag62}] {\includegraphics[trim = 2cm 0.4cm 2cm 1cm,clip,width=0.34\textwidth]{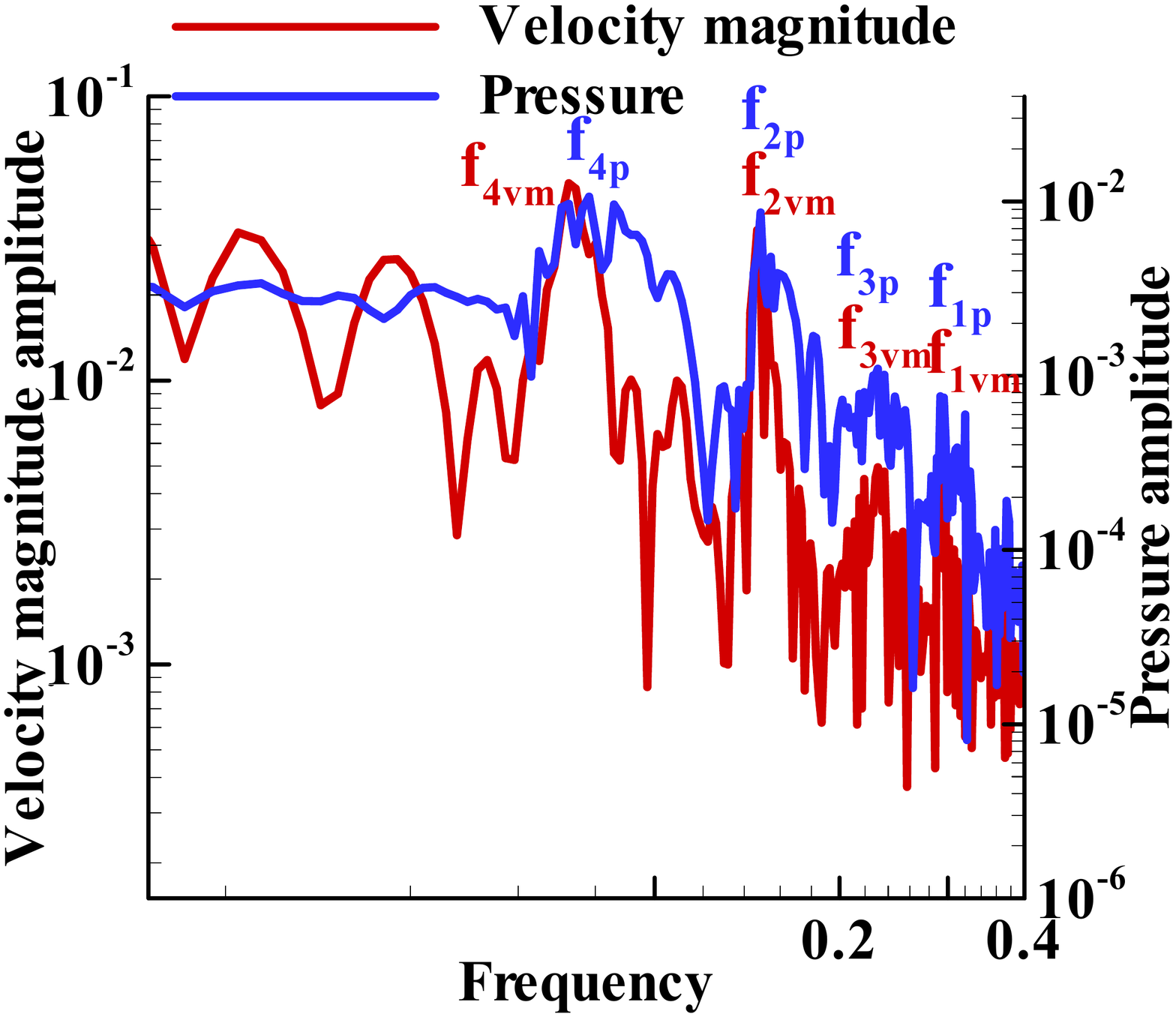}}}
\caption{FFT of Pressure and velocity magnitude at: a) ($x/D_{H}$,$y/D_{H}$) = (4,0), b) ($x/D_{H}$,$y/D_{H}$) = (22,0)  and c) ($x/D_{H}$,$y/D_{H}$) = (62,0)}
\label{p_vmag}
\end{figure*}
Since velocity is a vector, we compute its magnitude as $\sqrt{u^2+v^2}$ and the FFT of this quantity is plotted for all three regions behind the cylinder in figure \ref{p_vmag} as red-coloured lines. In the von-Karman region, the FFT of the velocity magnitude is similar to that of the streamwise velocity, but with only one single dominant frequency ($f_{1vm}$) as inferred from Fig. \ref{p_vmag4}. We previously surmised that two structures of $u$-velocity lie inside a single structure of $v$-velocity. Here, we add further to this hypothesis that the single structure of $v$-velocity consists of positive and negative structures of $v$, while the structures of $u$ are always positive. When taking the magnitude, these positive and negative $v$-structures will transform into two individual positive structures of $u$-velocity. As a result, the wavelength of the velocity magnitude structure becomes equal to that of the absolute value of $u$ -structure. This is the reason why $f_{1v}$ is not reflected in the velocity magnitude FFT. Fig. \ref{p_vmag} also shows that the velocity magnitude quantity of the von-Karman region is unaffected by the temporal wake development.

In the parallel vortex and secondary shedding regions, the FFT of velocity magnitude closely follows that of the streamwise velocity with four dominant frequencies (i.e. $f_{1vm}$, $f_{3vm}$, $f_{2vm}$, $f_{33vm}$) as seen Fig. \ref{p_vmag62}. This evidence suggests that the imprints of velocity magnitude is mostly determined by the streamwise velocity. This result may, sometimes, appear to be trivial after all the flow is predominately along the streamwise direction. Note, however, that: (i) the secondary low frequency of $u$-velocity in the von-Karman region is not reflected in the FFT of velocity magnitude, and (ii) $f_{4vm}$ appears in the secondary shedding region although it did not originate from the von-Karman region as with the case of the streamwise velocity. Therefore, these results are not completely trivial and the disappearance of $f_{2vm}$ in velocity magnitude FFT for the von-Karman region and its mysterious appearance in the secondary shedding region velocity magnitude FFT are further probed in section \ref{physical}.
\subsection{Pressure}
The final primitive variable of the flow filed is the pressure whose FFT is computed for all three regions and presented as blue-coloured lines in Fig. \ref{p_vmag}. Surprisingly, the FFT of pressure behaves similar as that of the streamwise velocity throughout the wake.

The pressure FFT of the von-Karman region has two dominant frequencies of $f_{1p}$ and $f_{2p}$ as seen in Fig. \ref{p_vmag4} (blue coloured lines). These frequencies (or the structures that have such frequencies) get convected downstream and the wake development causes two more additional frequencies in the parallel vortex region which, as a result, has four frequencies ($f_{1p}$, $f_{2p}$, $f_{3p}$ and $f_{4p}$) as noted in Fig. \ref{p_vmag22}. All these four frequencies survive in the secondary shedding region (see Fig. \ref{p_vmag62}). These observations are identical to that noted for the FFT of the streamwise velocity. 
\subsection{Temperature}
\begin{figure*}[htpb]
\centering
\mbox{\subfloat[\label{vt_4}] {\includegraphics[trim = 2cm 0.6cm 2cm 1cm,clip,width=0.33\textwidth]{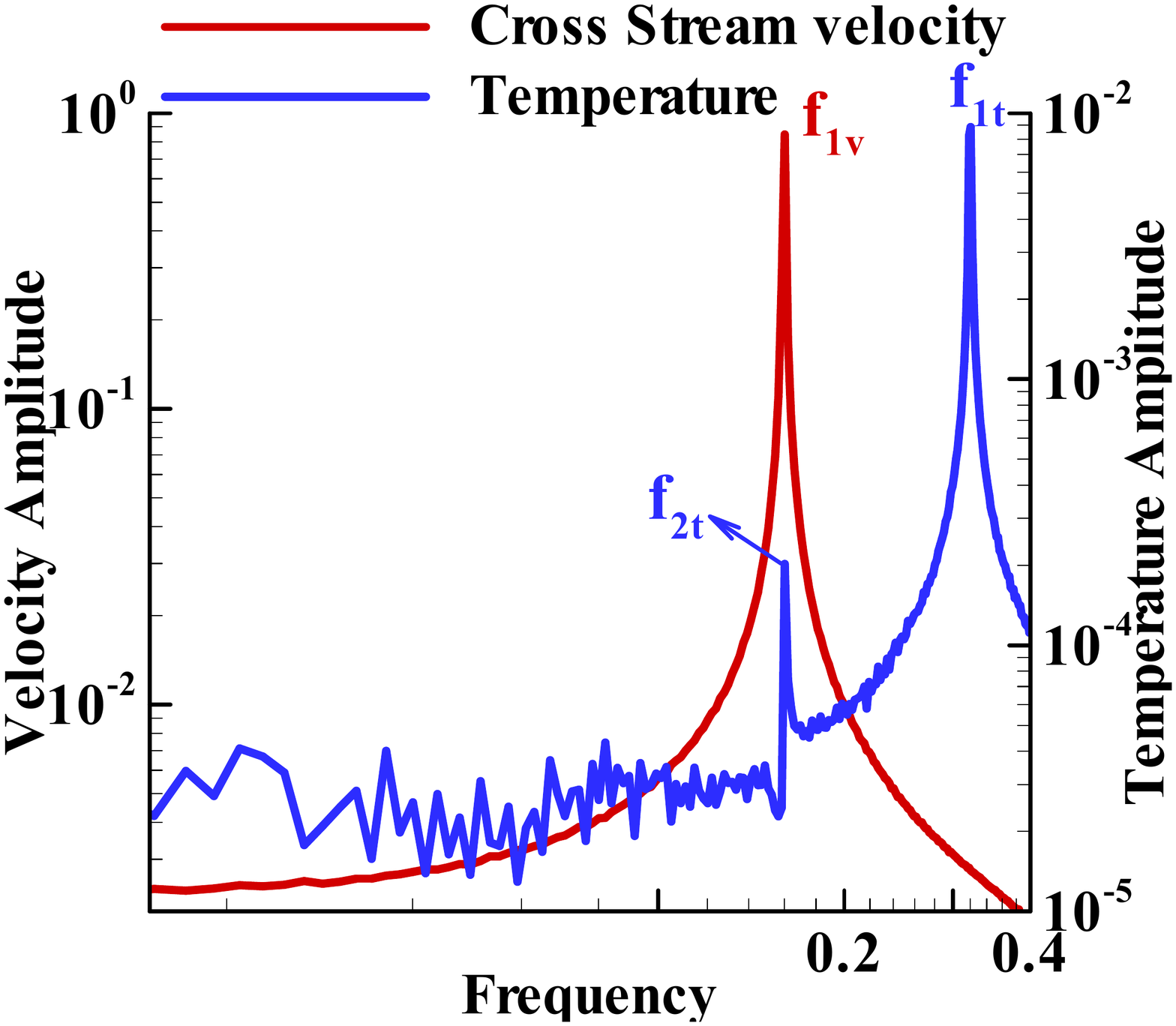}} \subfloat[\label{vt_22}] {\includegraphics[trim = 2cm 0.6cm 2cm 1cm,clip,width=0.33\textwidth]{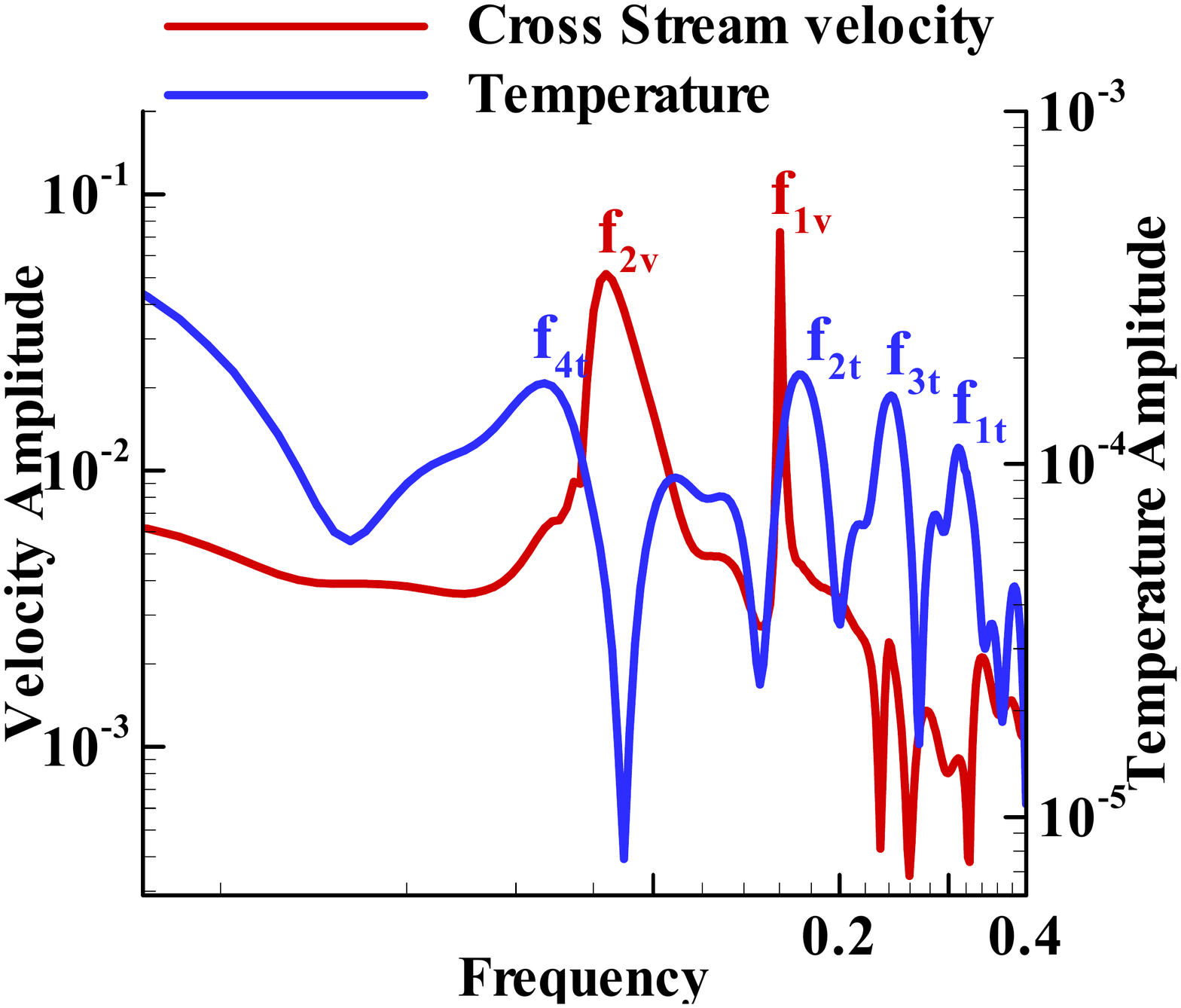}} \subfloat[\label{vt_62}] {\includegraphics[trim = 2cm 0.6cm 2cm 1cm,clip,width=0.33\textwidth]{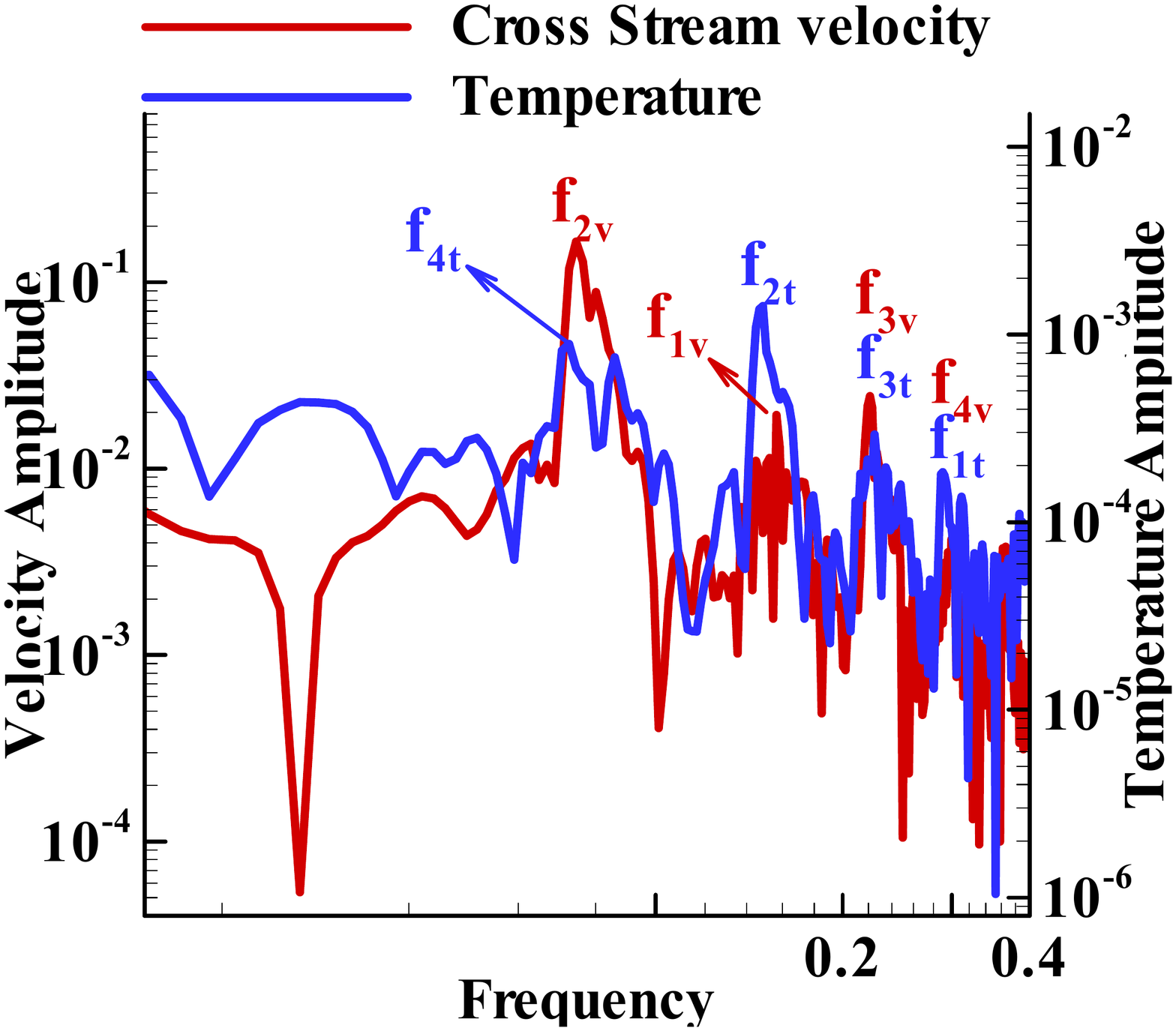}}}
\mbox{\subfloat[\label{ut_4}] {\includegraphics[trim = 2cm 0.6cm 2cm 1cm,clip,width=0.33\textwidth]{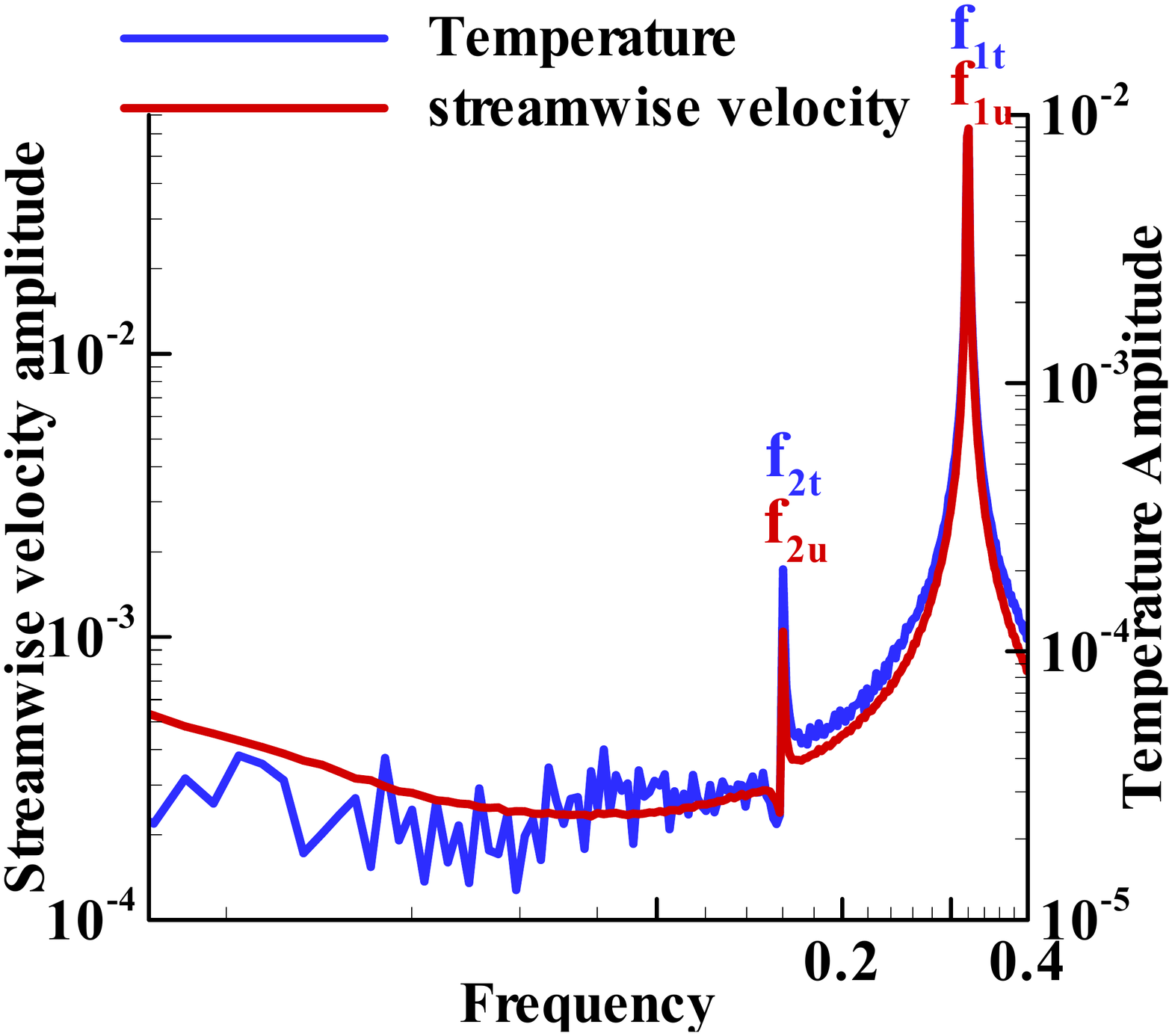}}\subfloat[\label{ut_22}] {\includegraphics[trim = 2cm 0.6cm 2cm 1cm,clip,width=0.33\textwidth]{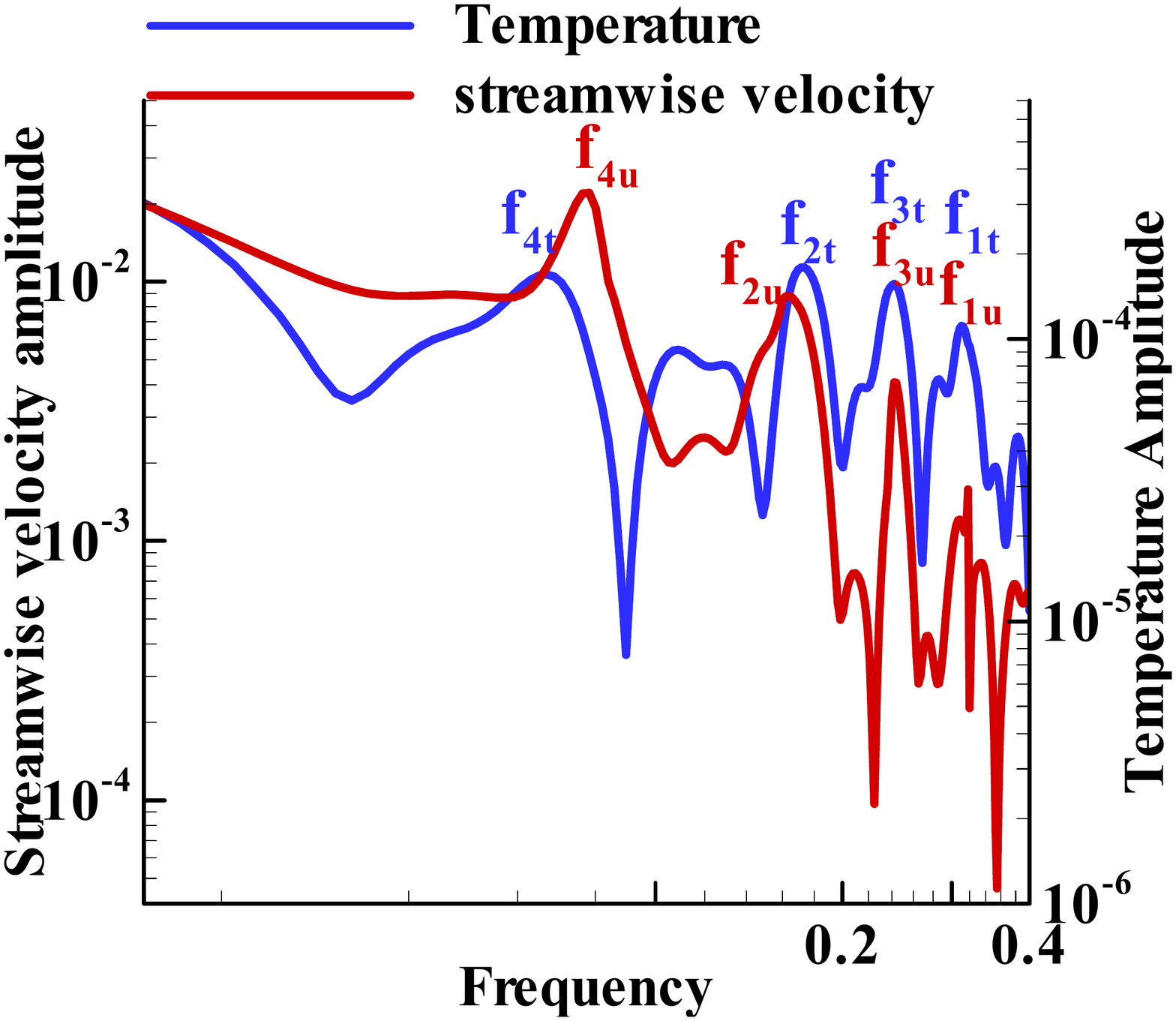}}\subfloat[\label{ut_62}] {\includegraphics[trim = 2cm 0.6cm 2cm 1cm,clip,width=0.33\textwidth]{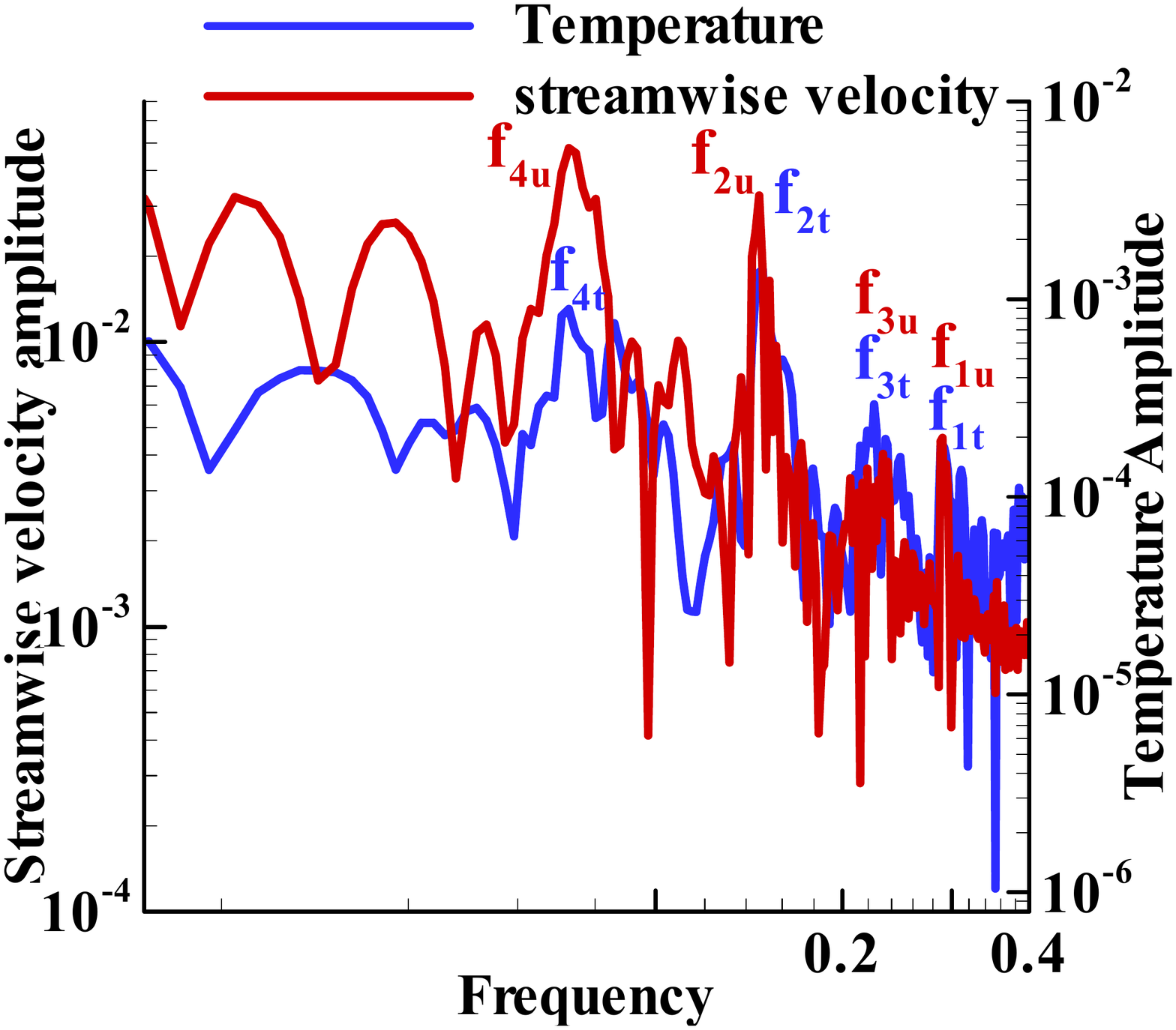}}}
\caption{FFT of cross stream velocity vs temperature (a-c) and FFT of streamwise velocity vs temperature (d-f) at ($x/D_{H}$,$y/D_{H}$) = (4,0), ($x/D_{H}$,$y/D_{H}$) = (22,0), and ($x/D_{H}$,$y/D_{H}$) = (62,0) respectively}
\label{vut}
\end{figure*}

The only primitive variable of the scalar field is the temperature. It is known from Fig. \ref{Inst_vort} that the temperature is trapped in vortices and the vortex shedding frequency is from the frequency of $v$-velocity as the Strouhal number computed from the lift coefficient signal is equal to $f_{1v}$. This observation tells us that the imprints of the temperature should be based on the cross-streamwise velocity because only the flow can influence the behaviour of the scalar and not the vice versa. However, the FFT of temperature reveals some interesting results.

The FFTs of the temperature at three regions are depicted in Fig. \ref{vut} along with the FFTs of the streamwise and cross-streamwise velocities. Figures \ref{vt_4} \& \ref{ut_4} show that the FFT of the temperature in the von-Karman region contains a primary larger frequency ($f_{1t}$) and a secondary low frequency, $f_{2t}$. Besides, the FFT of the temperature is comparable with that of the streamwise velocity, not with that of the cross-streamwise velocity.

The observation noted in the von-Karman region also holds true for the parallel vortex region as seen in Figs. \ref{vt_22} \& \ref{ut_22} that show four frequencies (i.e $f_{1t}$, $f_{2t}$, $f_{3t}$ and $f_{4t}$) in the FFT of the temperature. Here, $f_{1t}$ and $f_{2t}$ are due to the convective effect and $f_{3t}$ and $f_{4t}$ are due to the wake development.

Finally, Figs. \ref{vt_62} \& \ref{ut_62} reveal that the resonance of the temperature FFT with that of the cross-streamwise velocity survives even in the far-downstream secondary shedding region. In the secondary shedding region, all four frequencies from the parallel vortex region is retained.

We also know from the previous sub-section that the imprints of the streamwise velocity and the pressure are the same. Now, in this subsection, we have seen that the imprints of the temperature is based on the streamwise velocity. From these, we conclude that the scalar is predominately transported by the streamwise velocity and the pressure. This is contrary to the common belief that the scalar is transported by the vorticity and as a result the imprints of scalar should have been based on the cross-stream velocity.
\section{Spectral source of low frequency unsteadiness}\label{spectral}
In the previous section, we analyzed the primitive variables using FFT and noted the primary and secondary frequencies. In this section, we attempt to segregate the parts of the signal which cause the primary and secondary frequencies, thus we provide the spectral sources of these frequencies. In order to achieve this, we use an improved version of signal decomposition method introduced in \citet{pauls16}.
\subsection{Complex demodulation technique}
\begin{figure*}[htpb]
\centering
\mbox{\subfloat[\label{ampslo_4}] {\includegraphics[trim = 2.8cm 1.1cm 2.8cm 1cm,clip,width=0.33\textwidth]{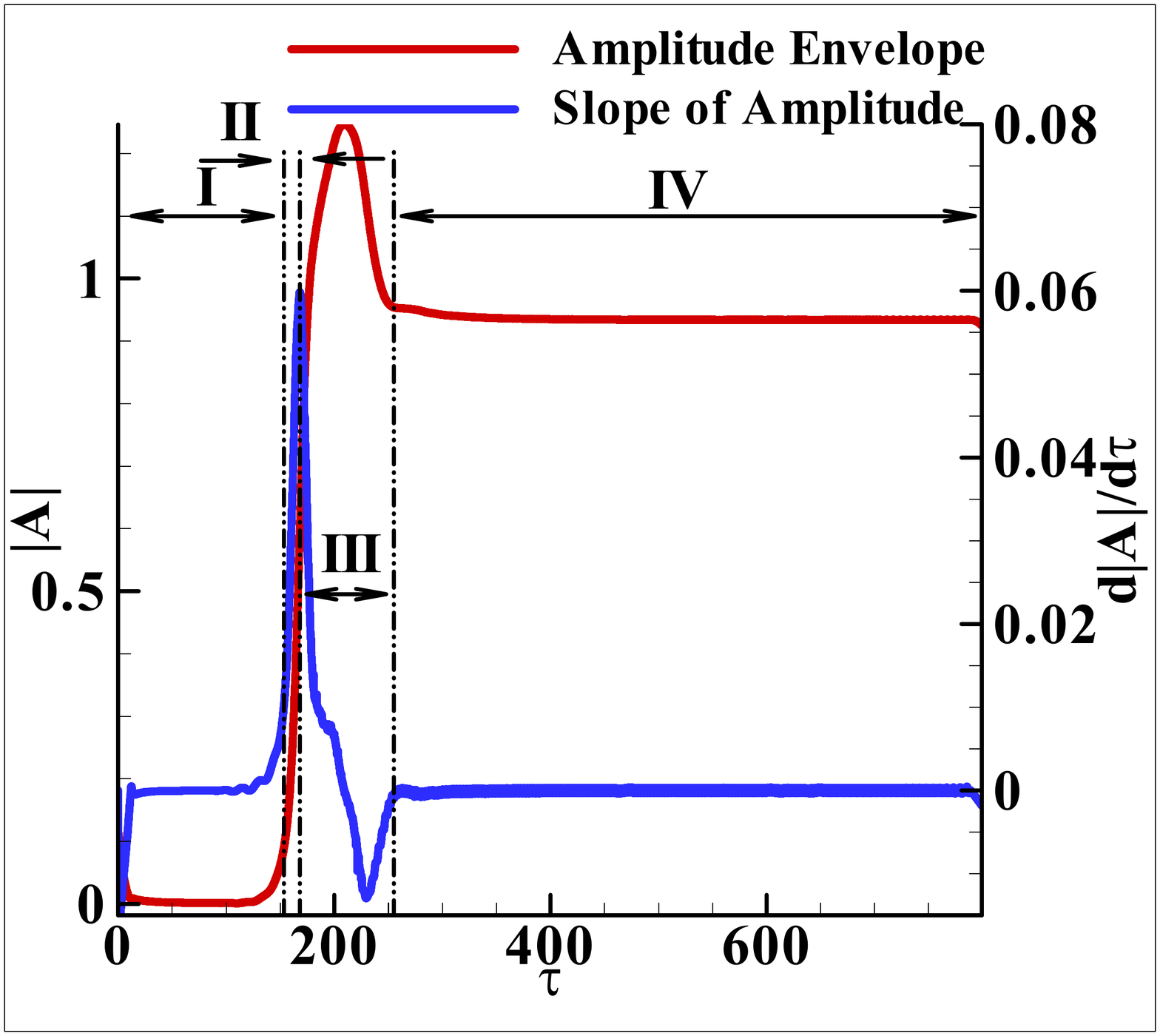}}\hspace{1mm} \subfloat[\label{ampslo_22}] {\includegraphics[trim = 3cm 1cm 1cm 1cm,clip,width=0.33\textwidth]{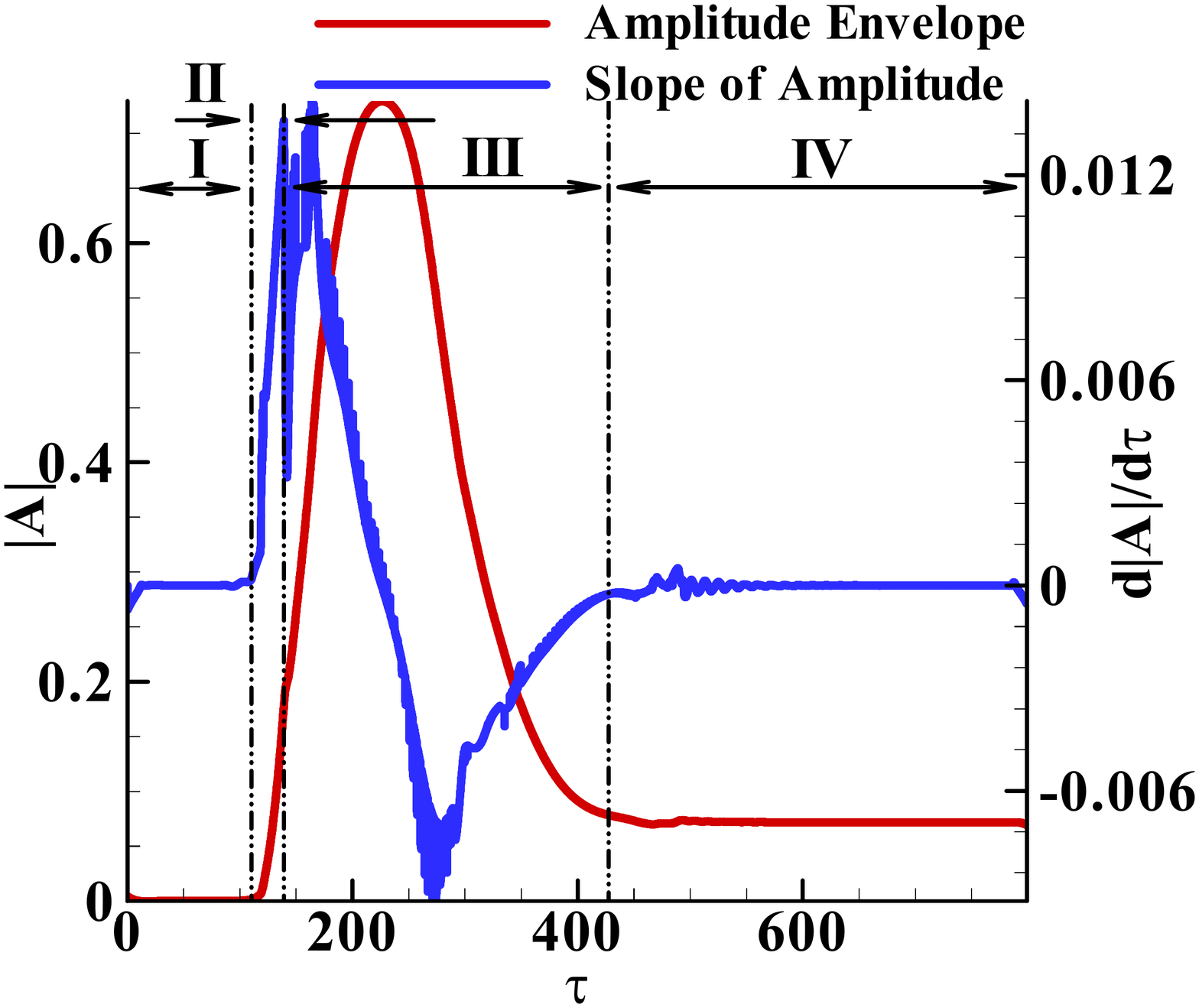}}\hspace{1mm}\subfloat[\label{ampslo_62}] {\includegraphics[trim = 2cm 1cm 2cm 1cm,clip,width=0.33\textwidth]{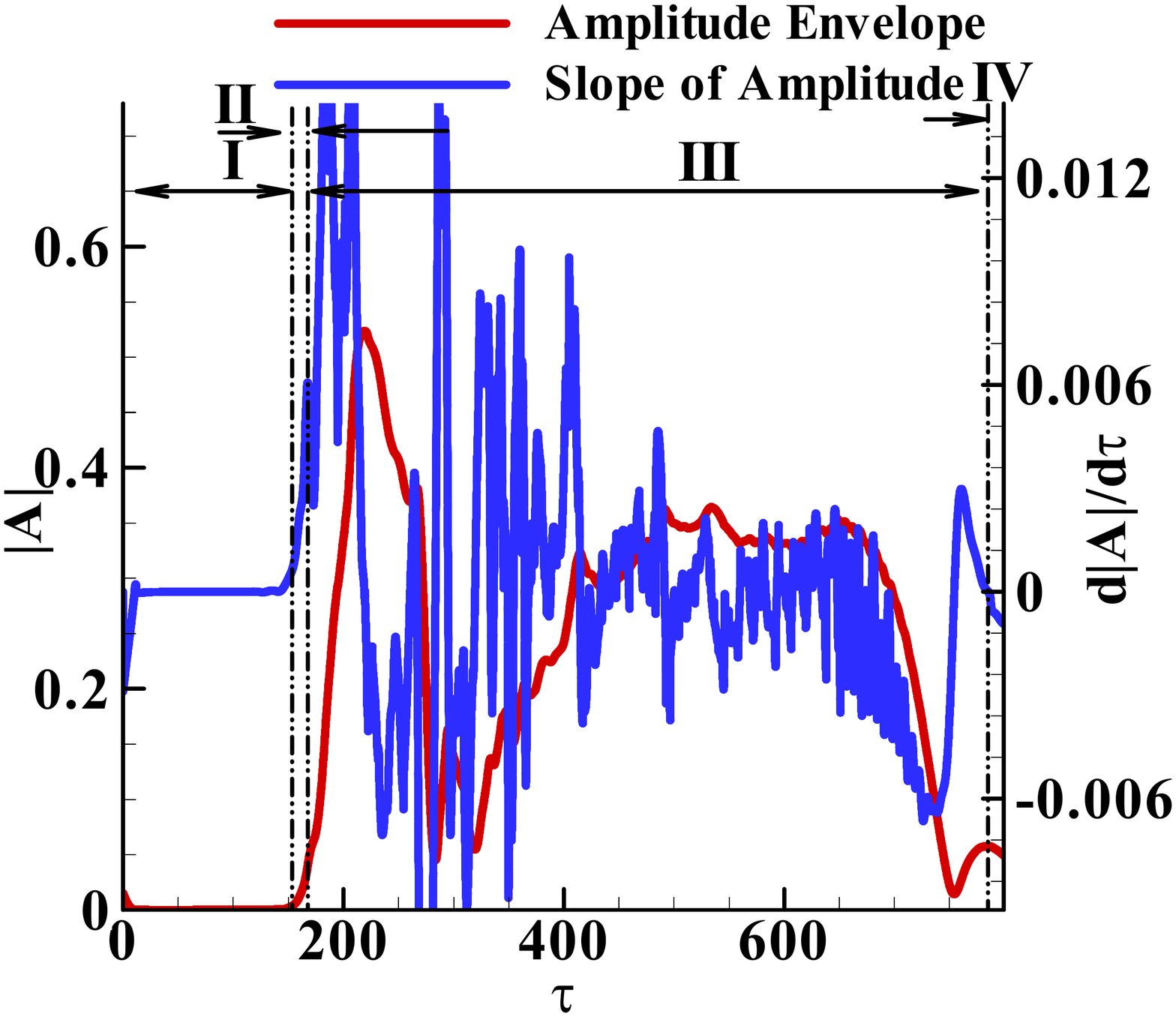}}}
\caption{Slope variations with time imposed on amplitude envelope at a) ($x/D_{H}$,$y/D_{H}$) = (4,0), b) ($x/D_{H}$,$y/D_{H}$) = (22,0)  and c) ($x/D_{H}$,$y/D_{H}$) = (62,0) demarcated into $I$ - Quasi steady regime, $II$- Linear growth regime, $III$ - Transition regime, and $IV$ - Saturation regime}. 
\label{ampslo}
\end{figure*}
The signal decomposition method we employ here is based on the complex demodulation technique (CDT) which gives the amplitude envelope of a signal as its output. The CDT consists of two steps. In the first stage of this method, a complex signal is generated. This complex signal has the actual signal as its real part and the Hilbert transform of the the actual signal as its imaginary part. In the second stage, the amplitude of the complex signal is computed which yields the amplitude envelope of the actual signal. \citet{pauls16} used the envelope signal to approximately segregate the different regions of the flow filed. In this study, we improve their method by separating different regimes in the signal based on the slope of the amplitude envelope. In this way, we are able to separate different regimes with an improved accuracy than that of \citet{pauls16}.

The aforementioned complex demodulation technique is applied to the cross stream velocity signal for the three regions and their corresponding amplitude envelopes are given in Fig. \ref{ampslo}. The slope of the amplitude envelope is used to decompose the signal into four different regimes as (i) Quasi steady ($I$) (ii) Linear growth ($II$) (iii) Transition ($III$) and (iv) Saturation regimes ($IV$). Fig. \ref{ampslo} also shows the demarcation of the time intervals of these regimes using the combined plots of the amplitude envelope and its slope.

The CDT presented in Fig. \ref{ampslo} paints the following picture regarding the temporal evolution of the flow. From $\tau$ = 0 to a certain time, the flow exhibits a quasi-steady state where its maximum unsteadiness is very low. This quasi-steady state is the equivalence of Figs. \ref{Inst_vort_a} \& \ref{Inst_vort_b} where only the standing vortex bubbles behind the cylinder can be seen. After this quasi-steady state, the disturbances in the flow grow linearly till the vortex shedding appears throughout the computational domain. The linear growth region can be better visualised in Fig. \ref{Inst_vort_c} which shows a partial shedding of the von-Karman vortex street. The regime of the signal where the shedding stabilizes is identified by a slow, yet positive $\left|A\right|$ v/s $\tau$ curve, and it is called the transition regime. Figures \ref{Inst_vort_d} and \ref{Inst_vort_e} represent this transition regime. After the transition regime, the disturbances reach a saturated state and the corresponding regime is called the saturation regime (see Fig. \ref{Inst_vort_f} for its physical representation).

Figures \ref{ampslo_4} and \ref{ampslo_22} show that all four regimes exits in the signal measured at the von-Karman and the parallel vortex regions. However, as seen in Fig. \ref{ampslo_62}, the secondary shedding region does not have a saturated region. This does not mean that the wake state has not reached a dynamic steady state, but it only means that the slope of the envelope does not become a constant in a region which is known for its chaotic nature.
\begin{table*}[htpb]
 \centering
 \begin{tabular}{|c|c|c|c|c|}
 \hline
 \textbf{($x/D_{H}$,$y/D_{H}$)} & \textbf{I} & \textbf{II} & \textbf{III} & \textbf{IV} \\ [0.6ex] 
\hline
 (4,0) & 0 - 153.2 & 153.2 - 168.1 & 168.1 - 255.2 & 255.2 - 799.1  \\[0.6ex]
 (10,0) & 0 - 103.1 & 103.1 - 157.4 & 157.4 - 287.8 & 287.8 - 799.1  \\[0.6ex]
 (22,0) & 0 - 110.1 & 110.1 - 139.2 & 139.2 - 427.7 & 427.7 - 799.1 \\[0.6ex]
 (38,0) & 0 - 123.8 & 123.8 - 168.1 & 168.1 - 685.1 & 685.1 - 799.1 \\[0.6ex]
 (50,0) & 0 - 128.0 & 128.0 - 155.3 & 155.3 - 741.0 & 741.0 - 799.1 \\[0.6ex]
 (82,0) & 0 - 153.3 & 153.3 - 167.6 & 167.6 - 784.9 & 784.9 - 799.1 \\[0.6ex]
 (76,0) & 0 - 164.6 & 164.6 - 181.9 & 181.9 - 799.1 & NA \\[0.5ex]
 (90,0) & 0 - 178.1 & 178.1 - 197.3 & 197.3 - 799.1 & NA \\[0.5ex]
 (96,0) & 0 - 181.8 & 181.8 - 203.9 & 203.9 - 799.1 & NA \\[0.5ex]
 (40,1) & 0 - 132.5 & 132.5 - 151.5 & 151.5 - 685.0 & 685.0 - 799.1 \\[0.6ex]
 (40,-1) & 0 - 136.8 & 136.8 - 172.2 & 172.2 - 700.7 & 700.7 - 799.1 \\[0.6ex]
 (60,1) & 0 - 158.9 & 158.9 - 167.6 & 167.6 - 799.1 & NA \\[0.6ex]
 (60,-1) & 0 - 150.8 & 150.8 - 162.4 & 162.4 - 799.1 & NA \\[0.6ex] 
\hline
\end{tabular}
\centering
\caption{Time Intervals of various regimes at various ($x/D_{H}$,$y/D_{H}$) locations (\textbf{I} - Quasi steady regime, \textbf{II}  - Linear growth regime, \textbf{III} - Transition regime and \textbf{IV} - Saturation regime), NA - Not Applicable}
\label{tab1}
\end{table*}
The time intervals of these four regimes at the probed locations are given in Table \ref{tab1}. Such time intervals can also be computed for any primitive variable although we have presented the result for the cross-stream velocity in Table \ref{tab1}.
\subsection{Cross-streamwise velocity}
\begin{figure*}[htpb]
\centering
\mbox{\subfloat[\label{vfft_2}] {\includegraphics[trim = 2.8cm 0.6cm 3cm 1cm,clip,width=0.33\textwidth]{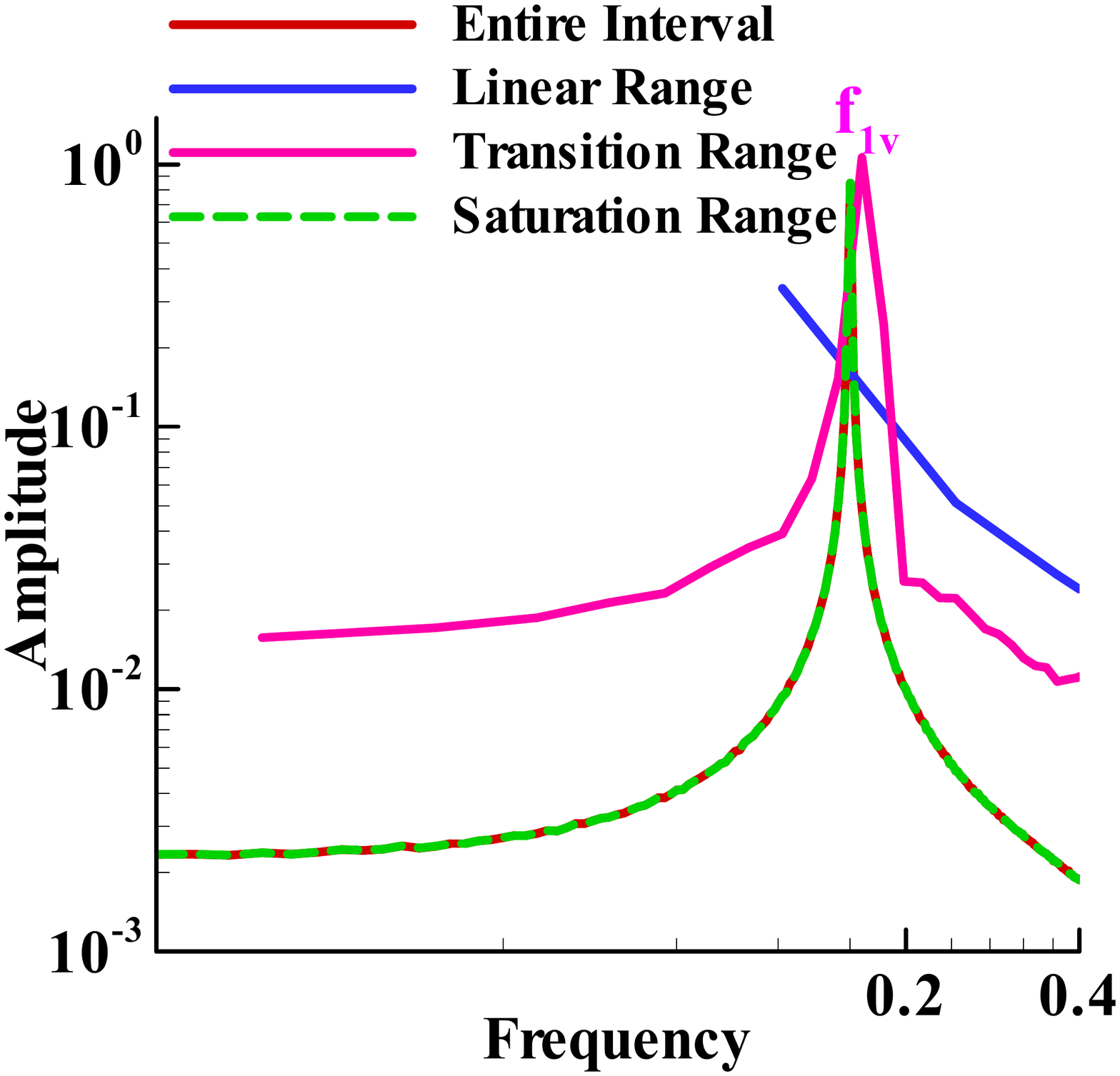}}\quad \subfloat[\label{vfft_22}] {\includegraphics[trim = 2.8cm 0.6cm 3cm 1cm,clip,width=0.33\textwidth]{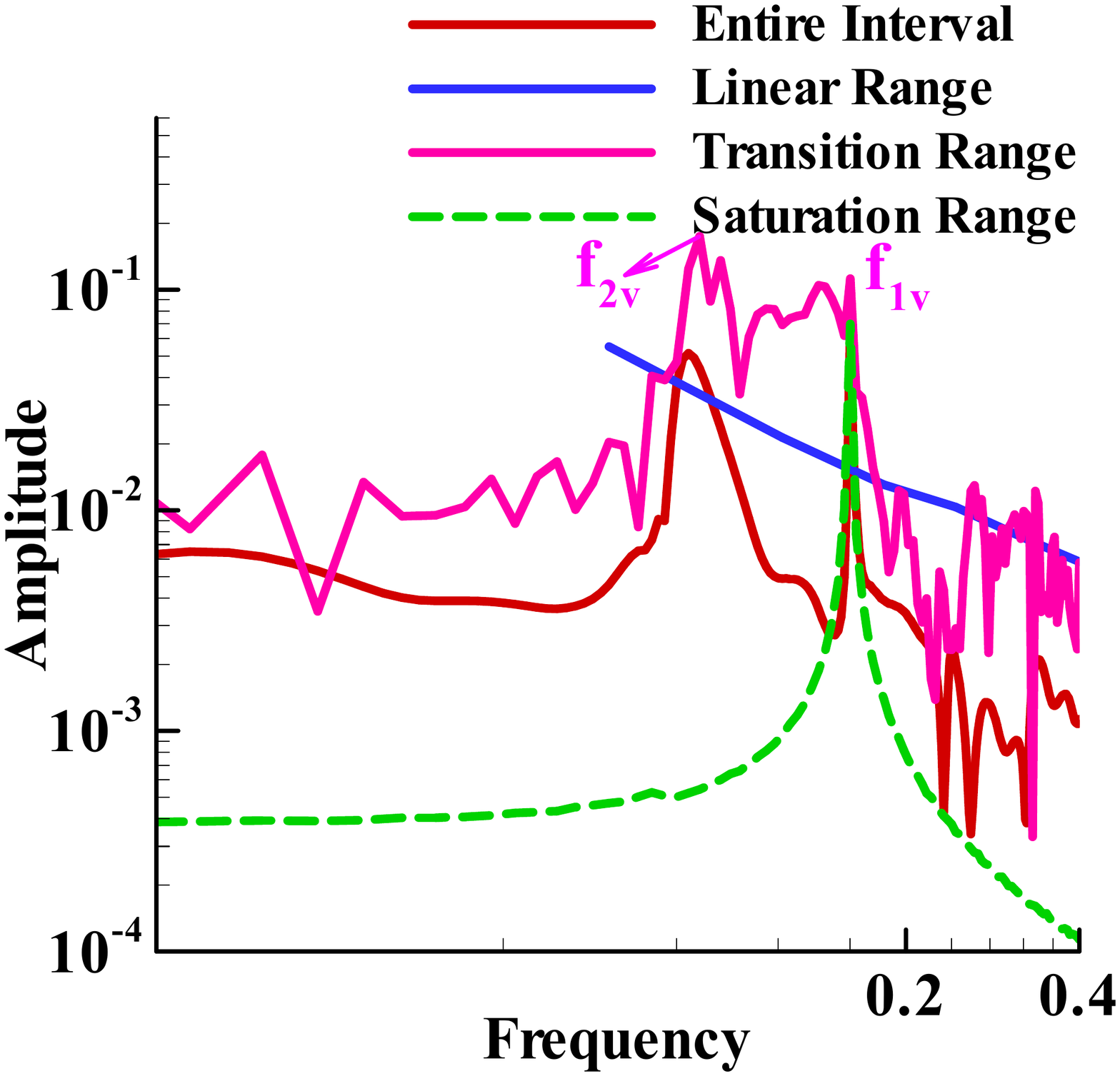}}\quad\subfloat[\label{vfft_62}] {\includegraphics[trim = 2.8cm 0.6cm 3cm 1cm,clip,width=0.33\textwidth]{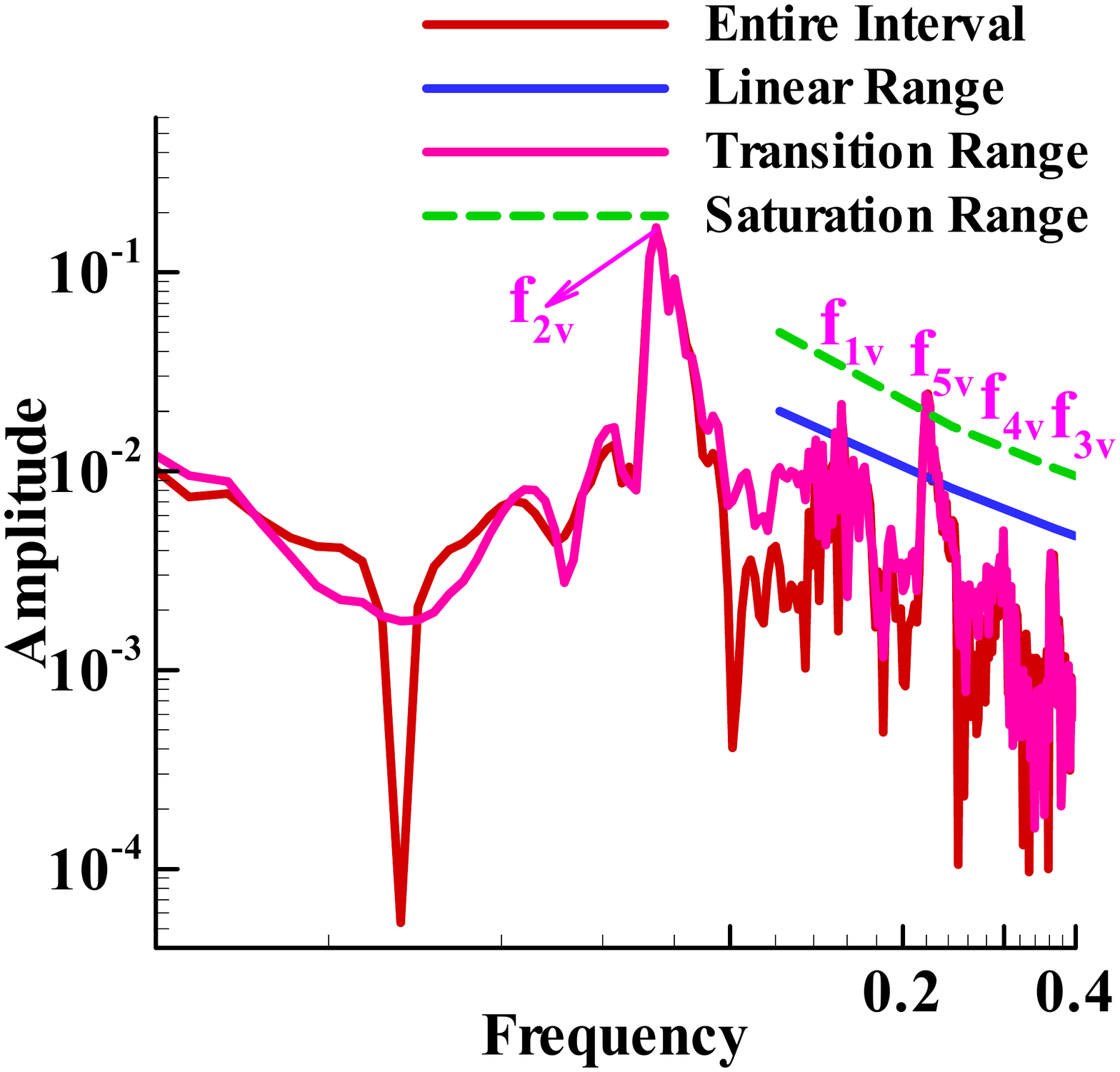}}}
\caption{FFT plots of cross stream velocity at a) ($x/D_{H}$,$y/D_{H}$) = (4,0), b) ($x/D_{H}$,$y/D_{H}$) = (22,0)  and c) ($x/D_{H}$,$y/D_{H}$) = (62,0)}
\label{vfft}
\end{figure*}

In this subsection, we decompose our cross-streamwise velocity signal into various segments according to the time intervals of various regimes reported in Table \ref{tab1}. Then we perform FFT for each of these segments. The result of this analysis is presented in Fig. \ref{vfft}. Since the quasi-steady regime part of the signal represents a steady flow, we did not consider FFT of it.

We know from our FFT analysis that the $v$-velocity structures have single frequency in the von-Karman region. Now, Fig. \ref{vfft_2} tells us that the single frequency ($f_{1v}$) is from the transition as well as the saturated regimes. This result further proves that the $v$-velocity in the von-Karman region is insensitive to transition or the wake development as transition regime generally contains flow structures of various wavelengths as seen in Fig. \ref{Inst_vort}.

Figure \ref{vfft_22} points out the source within the signal from which the primary and secondary frequencies originate for the parallel vortex region. When FFT is performed for the linear growth regime part of the signal, it does not result in any frequency even though some shedding occurs in this part of the signal (see the blue line in Fig. \ref{vfft_22}). When we take FFT of the transition regime portion of the signal, it has the imprint of the secondary low frequency ($f_{2v}$). Finally, the primary frequency ($f_{1v}$) is reflected only in the saturation part of the signal. Now we have obtained the exact locations within the signal from which the primary and secondary frequencies arise. In the case of the parallel vortex region, the part of the signal that falls under the transition region is responsible for the secondary low-frequency, and the part where the signal exhibits a saturated state is the spectral source of the primary frequency.

As we do not have a saturation regime for the secondary shedding region, the signals measured in this region is decomposed into three parts. Figure \ref{vfft_62} depicts one such example. However, in the secondary shedding region also, the linear growth regime of the signal does not have any frequency. On the other hand, the primary and secondary frequencies are embedded in the transition regime part of the signal. Thus, we have found the parts of the signal responsible for the primary and secondary frequencies in the FFT of $v$-velocity in all three regions.

\subsection{Streamwise velocity}
\begin{figure*}[htpb]
\centering
\mbox{\subfloat[\label{uvel_4}] {\includegraphics[trim = 2.8cm 0.6cm 3cm 1cm,clip,width=0.33\textwidth]{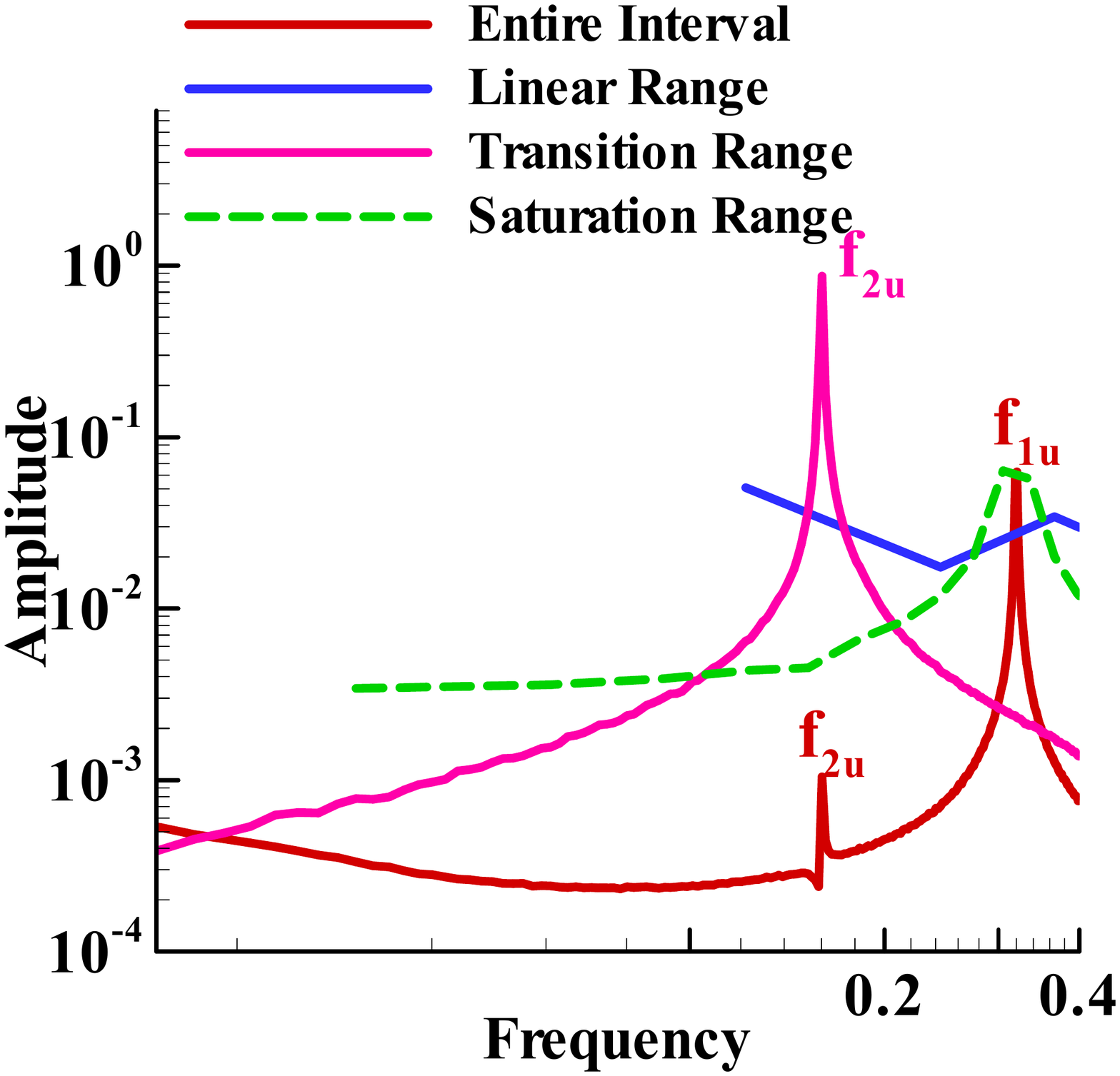}} \subfloat[\label{uvel_22}] {\includegraphics[trim = 2.8cm 0.6cm 3cm 1cm,clip,width=0.33\textwidth]{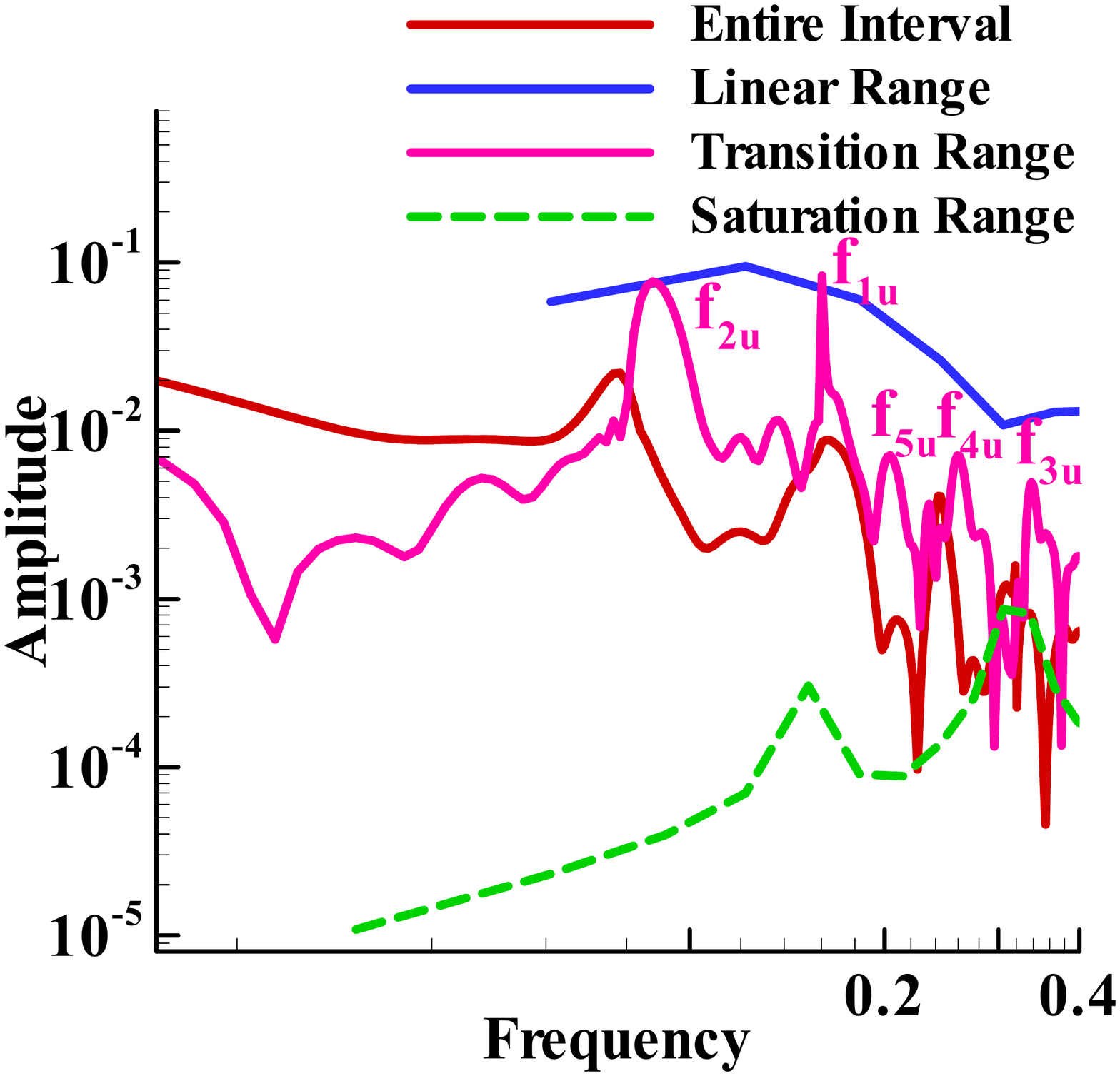}}\subfloat[\label{uvel_62}] {\includegraphics[trim = 2.8cm 0.6cm 3cm 1cm,clip,width=0.33\textwidth]{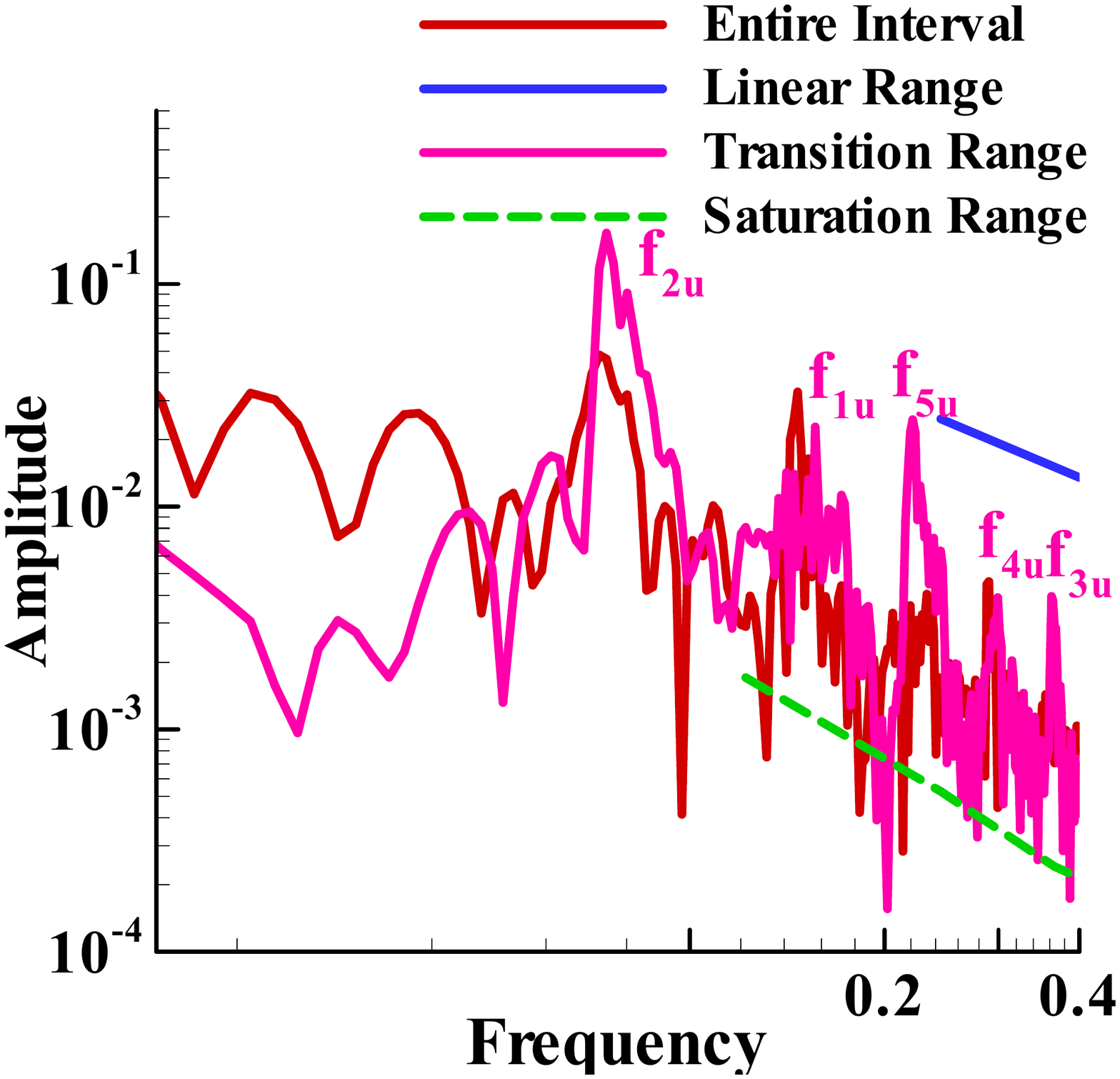}}}
\caption{FFT plots of streamwise velocity at a) ($x/D_{H}$,$y/D_{H}$) = (4,0), b) ($x/D_{H}$,$y/D_{H}$) = (22,0)  and c) ($x/D_{H}$,$y/D_{H}$) = (62,0)}
\label{uvel}
\end{figure*}

The procedure adopted in the previous section is now applied for the streamwise velocity, $u$. First, the time intervals of different segments of the signal is computed based on the complex demodulation technique explained earlier. Then we take FFT of each of the segment for a particular point for three regions and the result is presented in Fig. \ref{uvel}.

In the von-Karman region, Fig. \ref{uvel_4} shows that the primary frequency ($f_{1u}$), which is twice of the Strouhal frequency, is from the saturated regime of the signal. Interestingly, the $u$-velocity of the von-Karman region is sensitive to the wake development and results in a lower secondary frequency which is from the transition regime of the signal. It is indeed intriguing to note that the secondary low-frequency in the $u$ and $v$ FFTs of the von-Karman region always stem from the transition regime of the signal.

In the primary vortex and the secondary shedding regions, our analysis once again ascertains that the saturation regime is responsible for the primary frequency wherever it exists, and all the secondary frequencies are from the transitional regime of the signal. As we have shown earlier, the FFT of pressure is same as that of the streamwise velocity. Also, the imprints of the temperature FFT follows that of the $u$-velocity. Therefore, the discussion presented in this subsection is also applicable to FFTs of the pressure and the temperature.

\section{Physical source of low frequency unsteadiness}\label{physical}
In the previous section, we identified the part of the signal which is responsible for the primary and secondary frequencies. In this section, we take the instantaneous contours of flow and scalar field that correspond to that particular part of the signal in order to discover the physical source responsible for the primary and secondary frequencies.
\subsection{The von-Karman region}
\begin{figure*}[htpb]
\begin{center}
\mbox{\subfloat[\label{uvcont_2_180}] {\includegraphics[trim = 12 6 10 10,clip,width=0.19\textwidth]{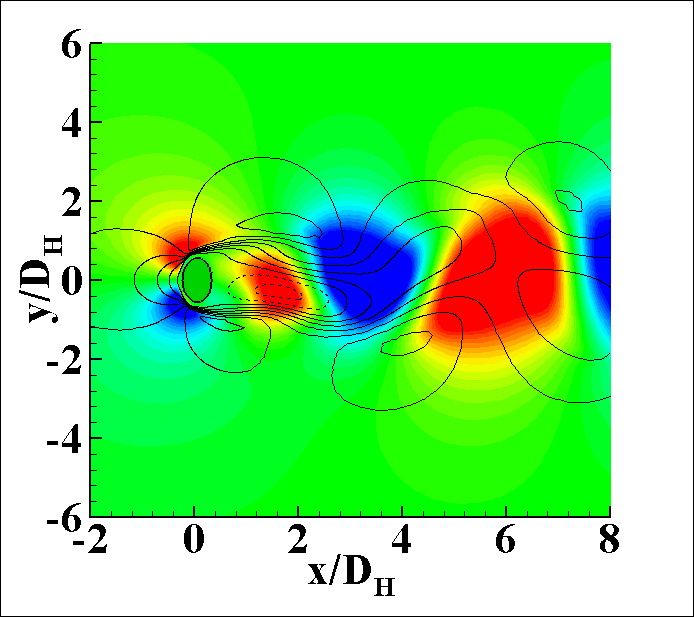}}\subfloat[\label{uvcont_2_223}] {\includegraphics[trim = 12 6 10 10,clip,width=0.19\textwidth]{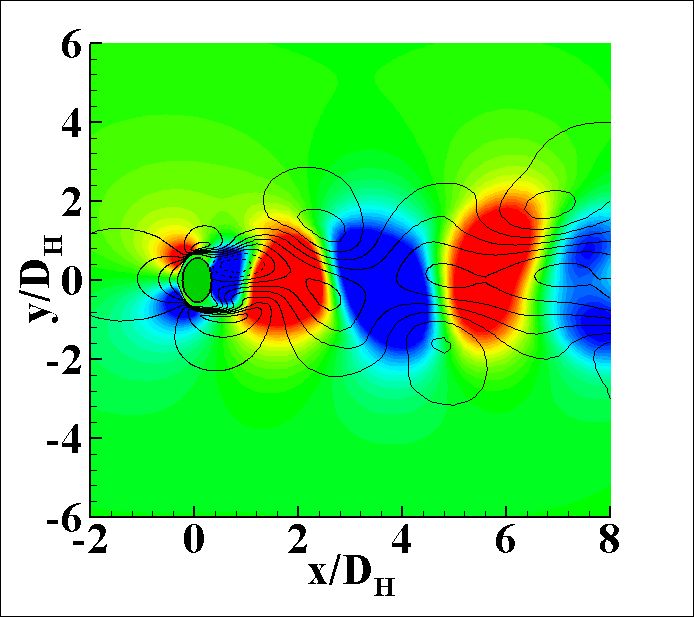}} \subfloat[\label{uvcont_2_404}] {\includegraphics[trim = 12 6 10 10,clip,width=0.19\textwidth]{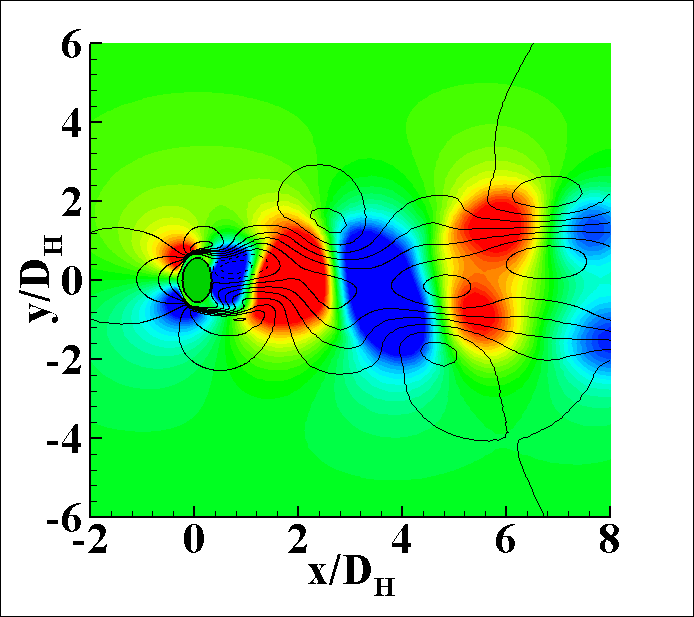}}\subfloat[\label{uvcont_2_460}] {\includegraphics[trim = 12 6 10 10,clip,width=0.19\textwidth]{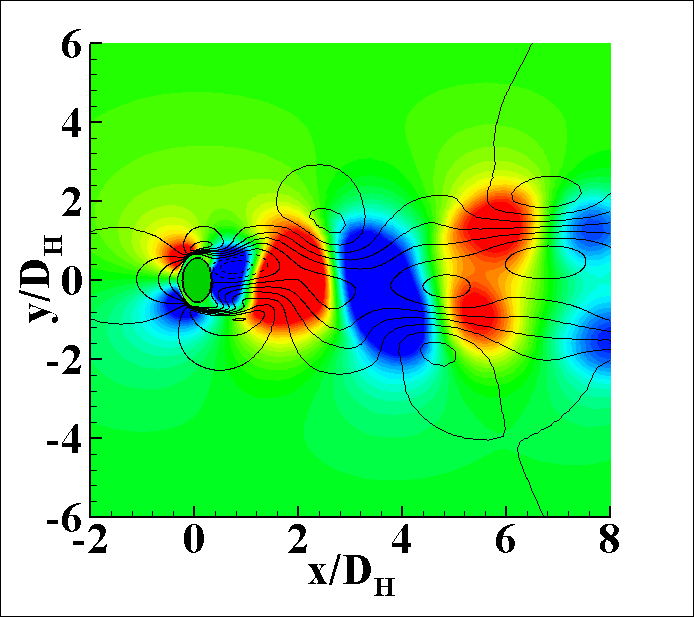}} \subfloat[\label{uvcont_2_740}] {\includegraphics[trim = 5 7 5 12,clip,width=0.23\textwidth]{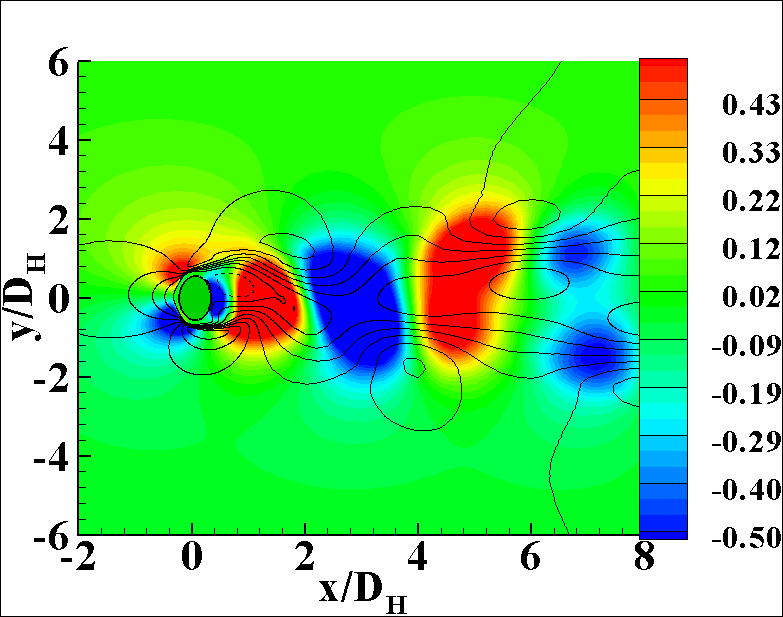}}}
\end{center}
\begin{center}
\mbox{\subfloat[\label{vort_ta}] {\includegraphics[trim = 12 7 10 10,clip,width=0.19\textwidth]{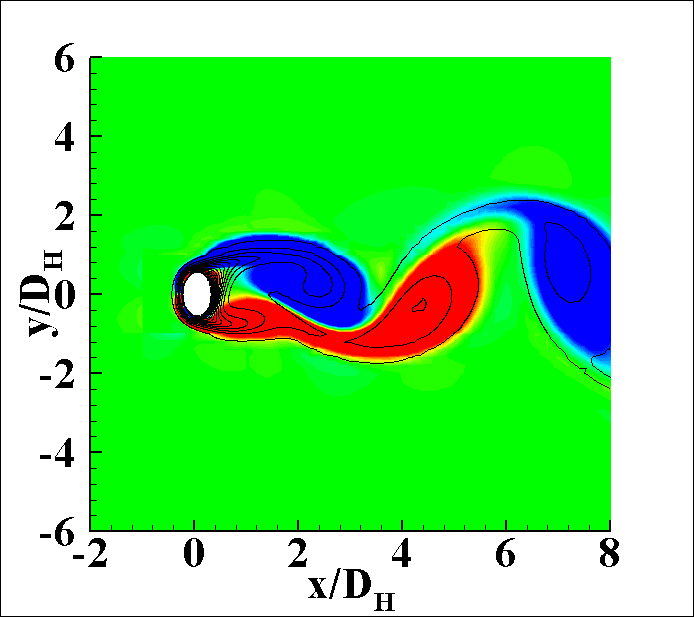}}\subfloat[\label{vort_tb}] {\includegraphics[trim = 12 4 10 10,clip,width=0.19\textwidth]{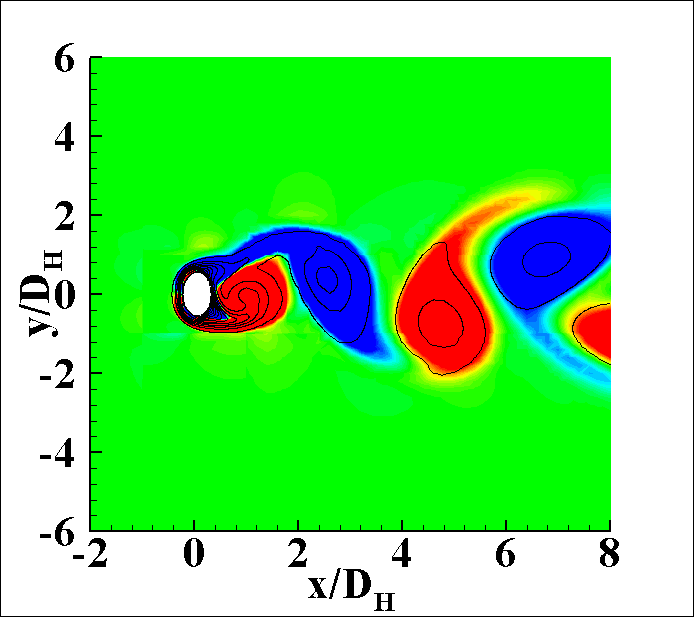}} \subfloat[\label{vort_tc}] {\includegraphics[trim = 12 5 10 10,clip,width=0.19\textwidth]{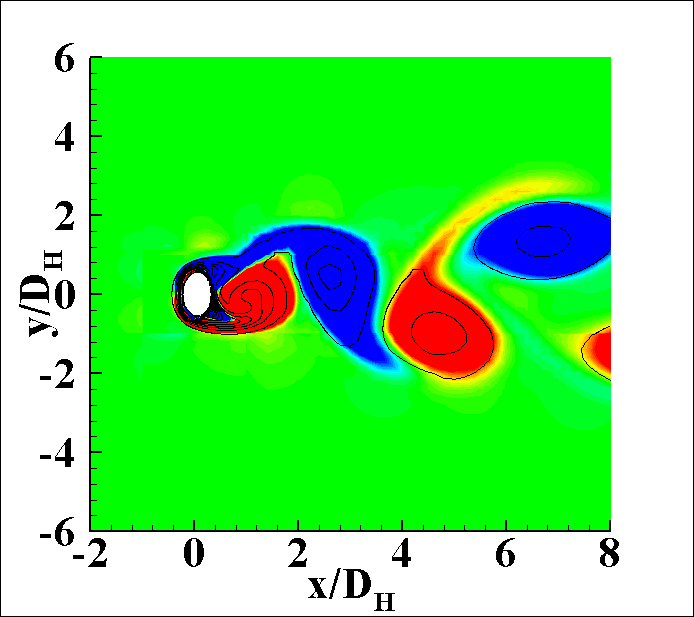}}\subfloat[\label{vort_td}] {\includegraphics[trim = 12 4 10 10,clip,width=0.19\textwidth]{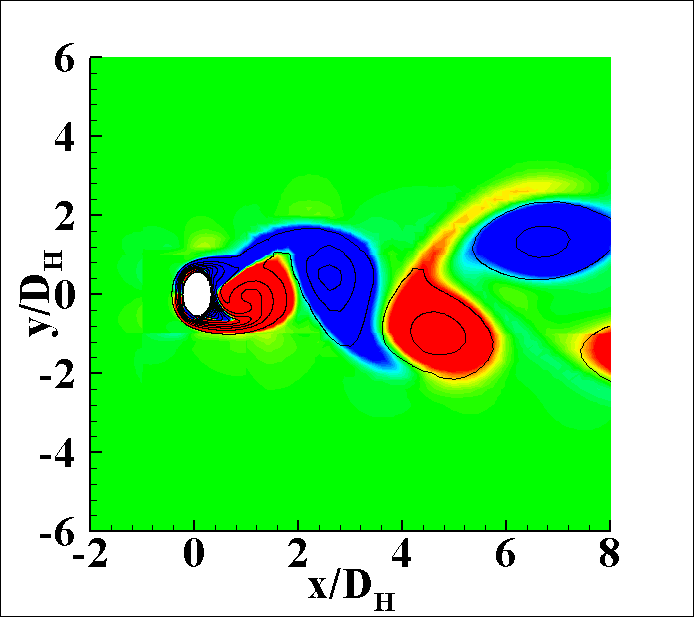}} \subfloat[\label{vort_te}] {\includegraphics[trim = 5 4 5 12,clip,width=0.23\textwidth]{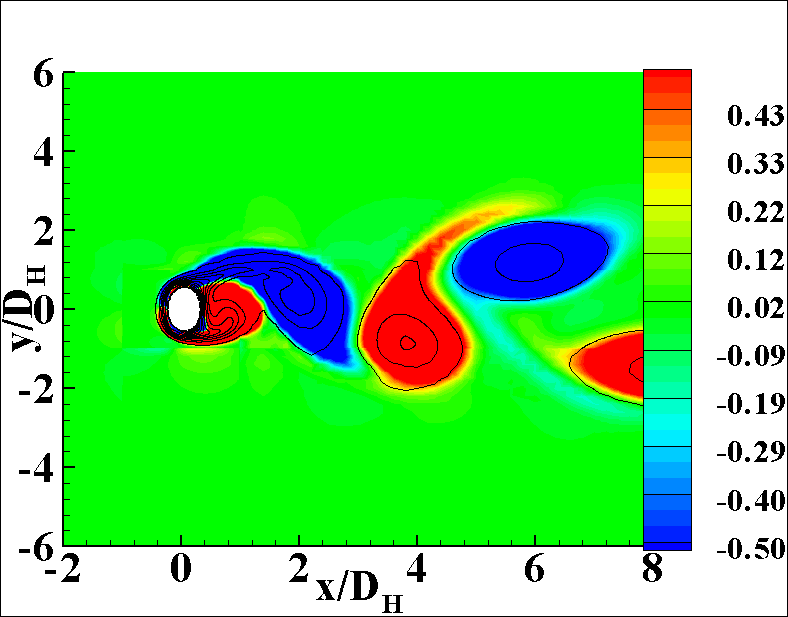}}}
\end{center}
\caption{ The temporal evolution of (a-e) u-contour lines (10 levels ranging from -0.4 to 1.6 (negative levels represented in dashed lines)) and v-contours (flood) and (f-j) are temperature contour lines(20 levels ranging from 0.01 to 0.6) at ($x/D_{H}$,$y/D_{H}$) = (4,0) respectively}
\label{uvcont}
\end{figure*}
We know from previous sections that the spectra of the streamwise velocity, the pressure and the temperature exhibit the primary and secondary frequencies while that of the cross-streamwise velocity results only one primary frequency in the von-Karman region of the wake. We also noted that the secondary low frequencies are from the transition regime of the wake development.

Figure \ref{uvcont} presents the temporal evolution of the flow in the von-Karman region in terms of streamwise, cross-streamwise, vorticity and temperature contours. The first two sub-figures (i.e. Figs. \ref{uvcont_2_180}, \ref{uvcont_2_223}, \ref{vort_ta} and \ref{vort_tb}) correspond to the transition regime of the signal and the last three (i.e. Figs. \ref{uvcont_2_404}-\ref{uvcont_2_740} and \ref{vort_tc}-\ref{vort_te}) are for the saturation regime.

The temporal evolution of $u$ and $v$ velocity structures of are given in Figs. \ref{uvcont_2_180}-\ref{uvcont_2_740}. From Fig. \ref{uvcont_2_180}, it can be clearly seen that the streamwise velocity structures are of positive values, while the structures of $v$-velocity have both positive and negative values. As a result, one structure of $v$-velocity contains a positive and negative structures of $v$-velocity, whereas one structure of $u$-velocity has just the contours of positive values. In the von-Karman region, all regimes of temporal development have two structures of $u$-velocity embedded inside one structure of $v$-velocity (see the dotted circles in Figs. \ref{uvcont_2_180}-\ref{uvcont_2_740}). This is the reason the frequency of the streamwise velocity structures is twice of that of the cross-streamwise velocity.

As seen in the Figs. \ref{uvcont_2_180}-\ref{uvcont_2_740}, the wavelength of one structure of $v$-velocity (i.e. both a positive and negative structure) does not change in all the figures although the size of the individual signed (i.e. positive or negative) structure changes from one region to the another. We have measured the wavelength of the $v$-velocity structure approximately as $4D_H$ and one can notice that this value does not change in all five Figs. \ref{uvcont_2_180}-\ref{uvcont_2_740} . This is the reason why the signal of $v$-velocity yielded only one single primary frequency for the von-Karman region. From the above explanation, we can see that the $v$-velocity of the von-Karman region is indeed affected by the temporal wake development (by means of having different sizes of individual signed structures) in contrast to the observation noted from the signal decomposition analysis. However, the overall effect nullifies this variation of wavelengths of signed structures of a single $v$-structure.

As noted earlier, the streamwise velocity structures are predominately of positive values. On the other hand, the temporal wake development makes these streamwise velocity structures susceptible to changing their wavelength much similar to the observation noted for the signed structures of $v$-velocity. We can note in Fig. \ref{uvcont_2_180} that the wavelength of $u$-velocity structure is approximately $2D_H$. This value transforms into almost half in the next Fig. \ref{uvcont_2_223}. This transmutation of size of the velocity structures causes the appearance of the lower frequency in the $u$-velocity signal of the von-Karman region.

As a summary from the discussions outlined above, we noted that the signed structures of velocity components in the von-Karman region undergo transmutation due to temporal wake development. However, the overall transmutation is nullified in the case of the cross-streamwise velocity as one structure of $v$-velocity consists of both positive and negative structures of cross-streamwise velocity. Since the spectra of pressure and streamwise velocity are commensurate, the explanation provided for the streamwise velocity is also applicable to the pressure.

The vorticity contours of the von-Karman region show that the vortices that are attached to the cylinder are stretched for a longer streamwise distance in the transition region (around $x/D_H \approx 3.5$) than that of the saturation region where they are stretched by less than $x/D_H \approx 1.5$. Due to this difference, the vortex structures that pass through $x/D_H=4$ posses diverse wavelengths. For instance, during the transition regime, structures of larger wavelength cross the von-Karman region as shown in Fig. \ref{vort_ta} and \ref{vort_tb}. On the other hand, structures of smaller wavelength traverse the von-Karman region during the saturation region (see Figs. \ref{vort_tc} - \ref{vort_te}). This observation validates our previous result from the signal decomposition that lower frequencies arise from the transition regime. Since the passive scalar is trapped in vorticity, the spectra of scalar exhibit the behaviour of vorticity. This is the reason the spectrum of temperature in the von-Karman region showed the presence of larger primary frequency and a secondary low frequency.
\subsection{The parallel vortex region}
\begin{figure*}[htpb]
\begin{center}
\mbox{\subfloat[\label{uvcont_11_180}] {\includegraphics[trim = 12 7 10 10,clip,width=0.19\textwidth]{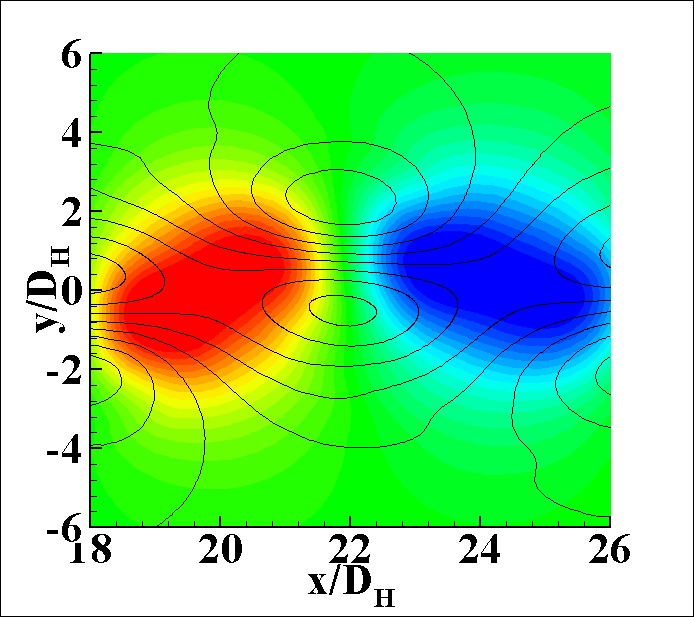}}\subfloat[\label{uvcont_11_223}] {\includegraphics[trim = 12 7 10 10,clip,width=0.19\textwidth]{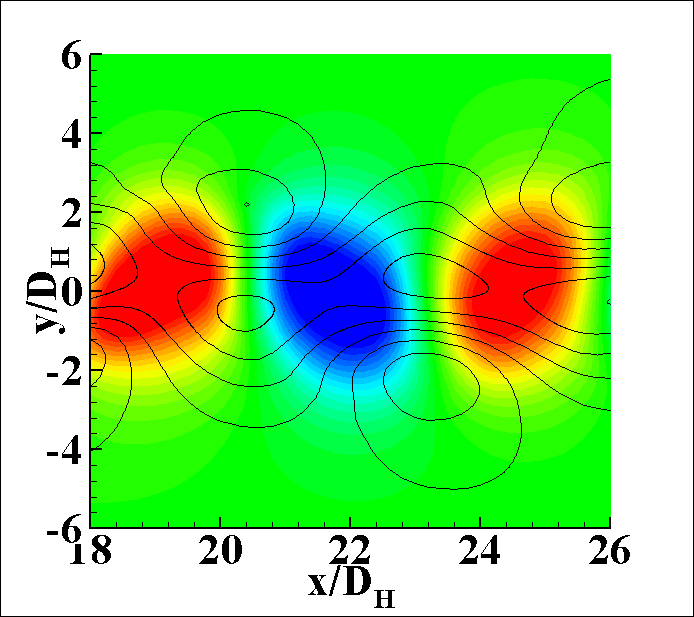}} \subfloat[\label{uvcont_11_404}] {\includegraphics[trim = 12 7 10 10,clip,width=0.19\textwidth]{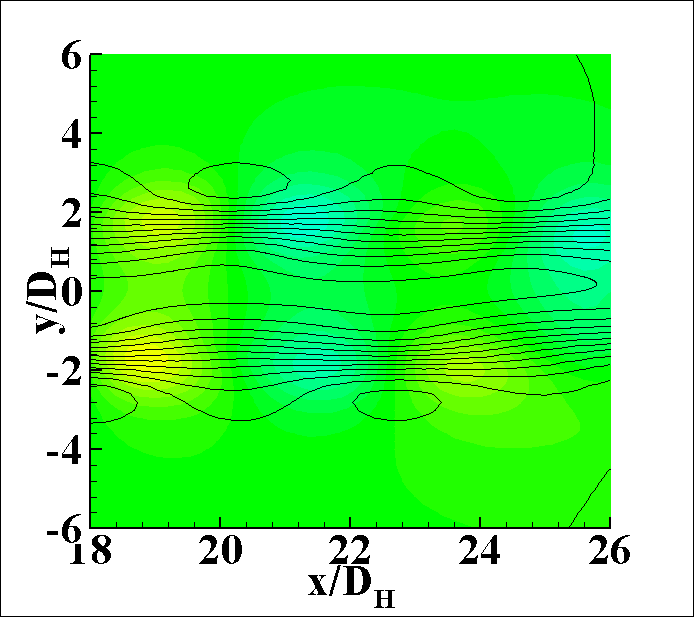}}\subfloat[\label{uvcont_11_460}] {\includegraphics[trim = 12 7 10 10,clip,width=0.19\textwidth]{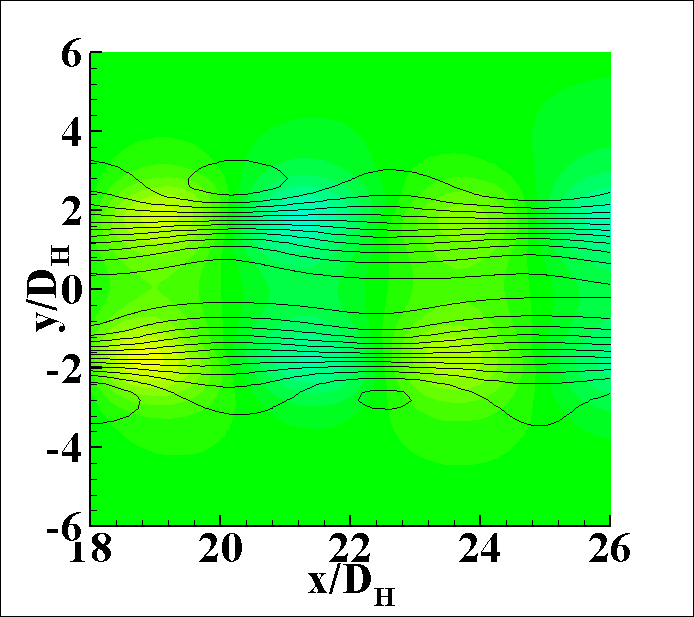}} \subfloat[\label{uvcont_11_740}] {\includegraphics[trim = 4 8 5 12,clip,width=0.21\textwidth]{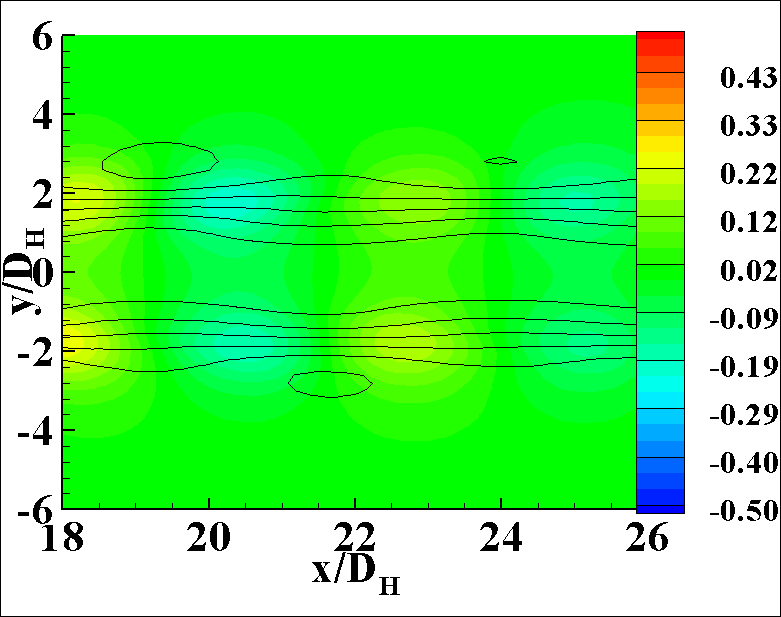}}}
\end{center}
\begin{center}
\mbox{\subfloat[\label{vort_tf}] {\includegraphics[trim = 12 8 10 10,clip,width=0.19\textwidth]{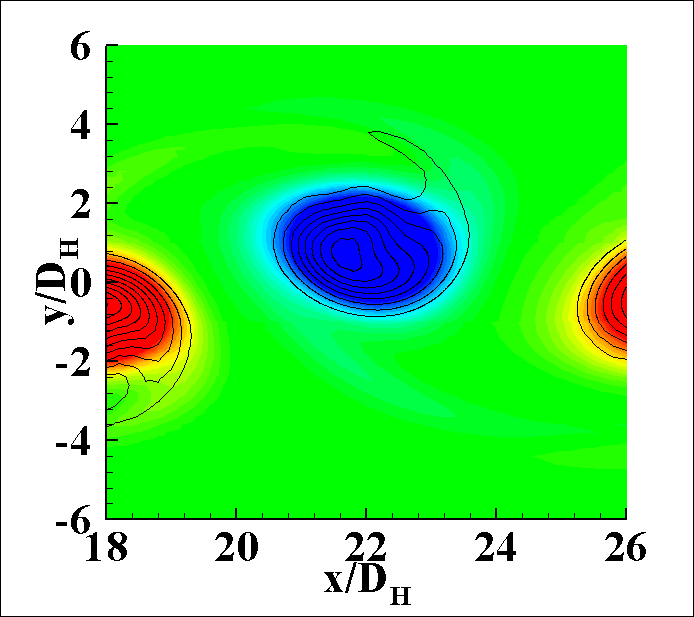}}\subfloat[\label{vort_tg}] {\includegraphics[trim = 12 8 10 10,clip,width=0.19\textwidth]{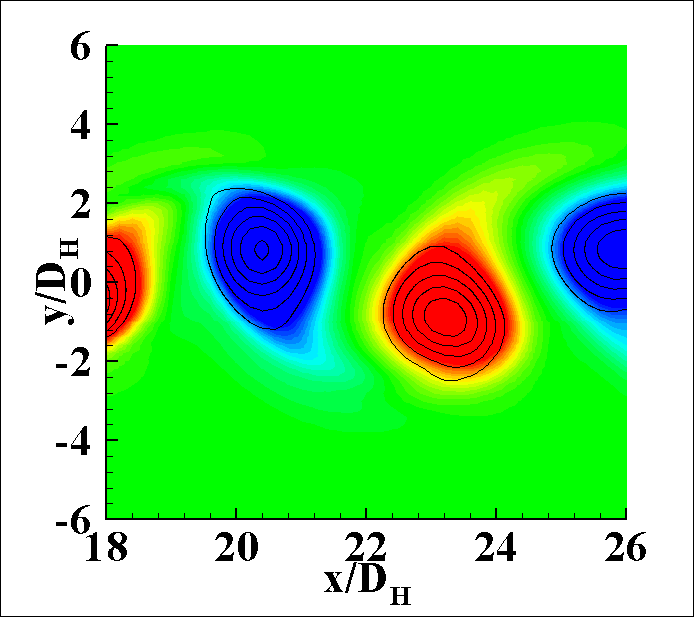}} \subfloat[\label{vort_th}] {\includegraphics[trim = 12 8 10 10,clip,width=0.19\textwidth]{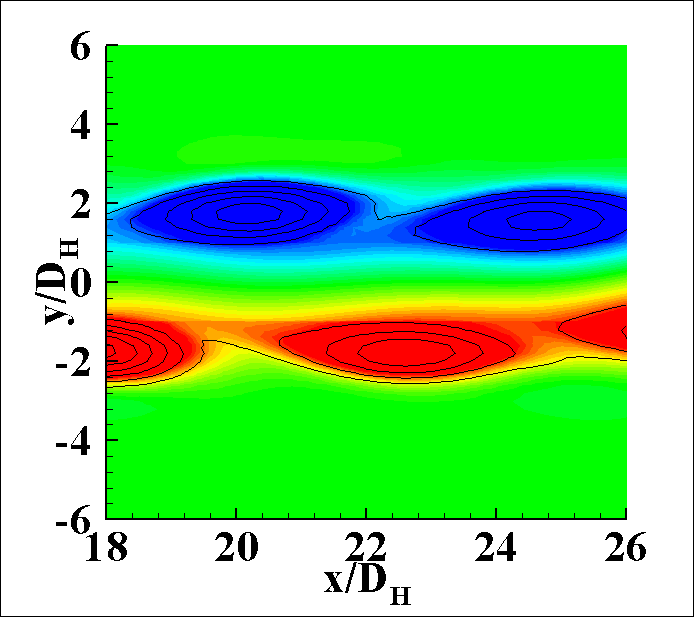}}\subfloat[\label{vort_ti}] {\includegraphics[trim = 12 8 10 10,clip,width=0.19\textwidth]{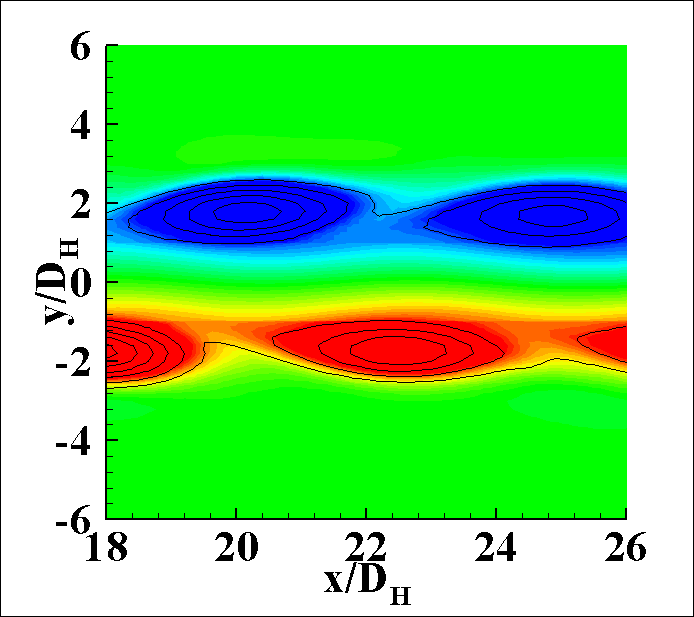}} \subfloat[\label{vort_tj}] {\includegraphics[trim = 5 7.5 5 12,clip,width=0.22\textwidth]{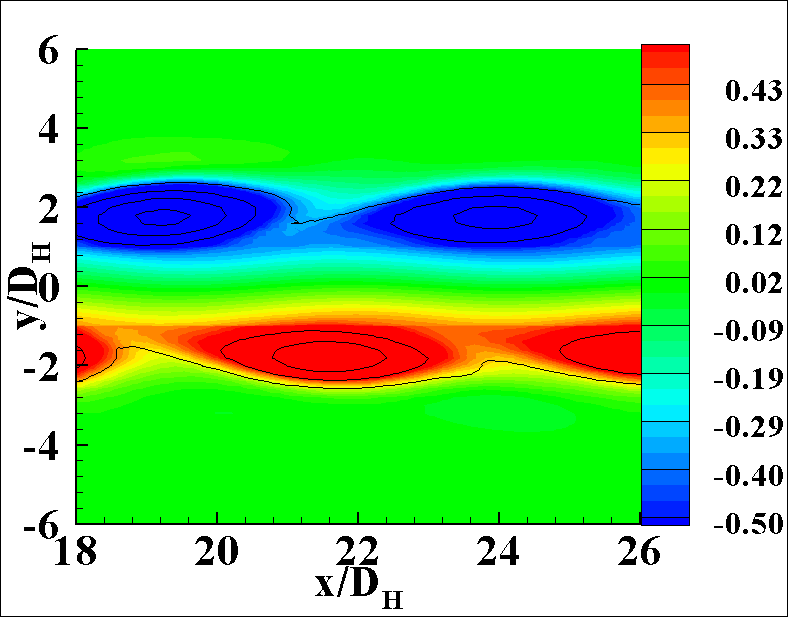}}}
\end{center}
\caption{The temporal evolution of (a-e) u-contour lines (10 levels ranging from -0.4 to 1.6 (negative levels represented in dashed lines)) and v(flood) contours and (f-j) are temperature(20 levels ranging from 0.01 to 0.6) contour lines at ($x/D_{H}$,$y/D_{H}$) = (22,0) respectively}
\label{uvcont2}
\end{figure*}

The signal decomposition revealed that the primary and secondary frequencies appear in all primitive variables of the parallel vortex region. Figure \ref{uvcont2} portrays the temporal evolution of the momentum and thermal wakes in the parallel vortex region (i.e. $18\leq x/D_H \leq 26$). As with the previous figure, here also, the first two subplots (i.e. Figs. \ref{uvcont_11_180}, \ref{uvcont_11_223}, \ref{vort_tf} and \ref{vort_tg}) represent the transition regime while the other three are for the saturated regime of the temporal wake development. The $v$-velocity structures show the presence of periodic structures during the transition regime (see Figs. \ref{uvcont_11_180} and \ref{uvcont_11_223}), while these structures align parallel to each other in the saturation regime as seen in Figs. \ref{uvcont_11_404}-\ref{uvcont_11_740}. This observation reflects that the primary von-Karman vortex street of the transition regime undergoes a transformation and it obtains a parallel row of vortices in the saturated state of the wake development. This can also be noted in the vorticity contours of Figs. \ref{vort_tf}-\ref{vort_tj}.

The wavelengths of the $v$-velocity structures change due to temporal wake development. This can be easily seen by comparing Figs. \ref{uvcont_11_180} and \ref{uvcont_11_223}. We have only one (i.e. two signed) $v$-velocity structure in Fig. \ref{uvcont_11_180} whose wavelength is approximately $8x/D_H$. At the same time, there exists more than one $v$-velocity structure for the same region in Fig. \ref{uvcont_11_223} along with the change of wavelength which is approximately computed as $5x/D_H$. This is completely different from the behaviour noted for the von-Karman region where the wavelength of the complete $v$-velocity structure (which contains two signed $v$-structures) did not change although the wavelengths of signed $v$-structures got modified. This observation is the direct impact of the temporal wake development on the $v$-velocity structures of the parallel vortex region. The larger $v$-velocity structures that are reminiscent of the large wavelength structures result in low secondary frequency in the spectra of the transition regime's $v$-velocity. The large primary frequency is due to the structures that are smaller than the large wavelength structures as seen in Figs. \ref{uvcont_11_404}-\ref{uvcont_11_740}. 

Concerning the streamwise velocity structures, they also undergo change in their wavelength ranging from $4x/D_H$ (see Fig. \ref{uvcont_11_180} and again note that it is half of the wavelength of the $v$-velocity structure) to $2x/D_H$ in Fig. \ref{uvcont_11_223} in the transition regime. Apart from these two wavelengths, we can also see $u$-velocity structures of various sizes and they get reflected as smaller frequencies lower than that of the primary frequency in the spectra of the transition part of the $u$-velocity. As noted for the $v$-velocity, the $u$-velocity structures of the saturation regime (Figs. \ref{uvcont_11_404}-\ref{uvcont_11_740}) also do not change their wavelengths and they result in a single frequency in the FFT of saturated $u$-velocity as we noted in Fig. \ref{uvel_22}. 

Finally, the vorticity contours shown in Figs. \ref{vort_tf}-\ref{vort_tj} also reflect the scenario explained in the above paragraphs. That is, the wavelengths of vortex structures get modified in the transition regime (see Figs. \ref{vort_tf} \& \ref{vort_tg}) and they remain the same size throughout the saturation regime. As the scalar is trapped in the vortical structures, the largest frequency in the scalar spectra of the parallel vortex region is from the saturated state of the wake development, while all other smaller frequencies of the spectra are due to the transitional state of the wake development.
\subsection{The secondary shedding region}
\begin{figure*}[htpb]
\begin{center}
\mbox{\subfloat[\label{uvcont_31_180}] {\includegraphics[trim = 6 4 10 10,clip,width=0.19\textwidth]{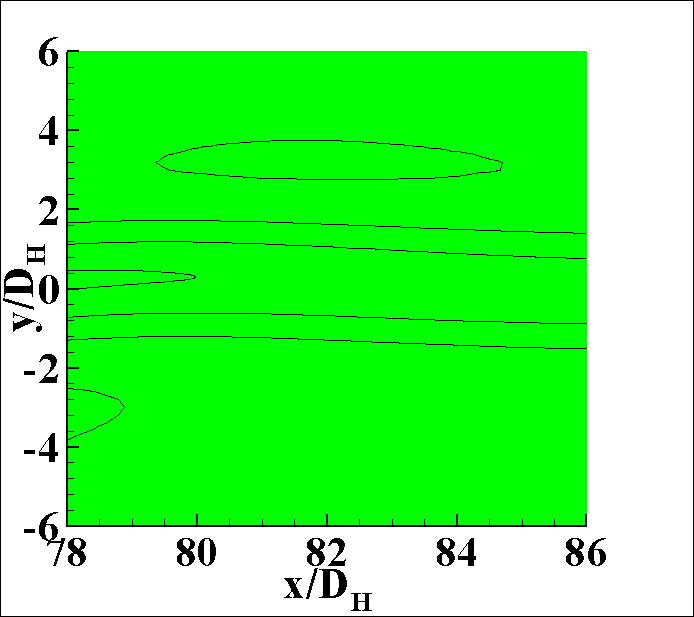}}\subfloat[\label{uvcont_31_223}] {\includegraphics[trim = 6 4 10 10,clip,width=0.19\textwidth]{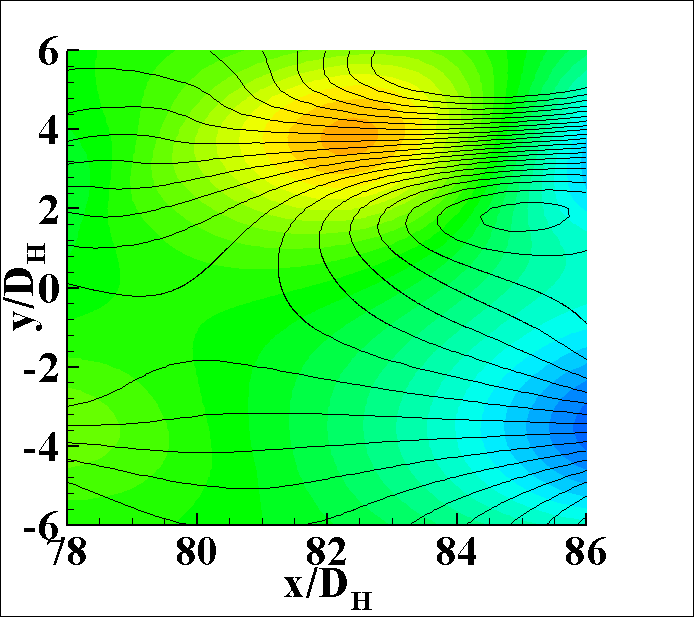}} \subfloat[\label{uvcont_31_404}] {\includegraphics[trim = 6 4 10 10,clip,width=0.19\textwidth]{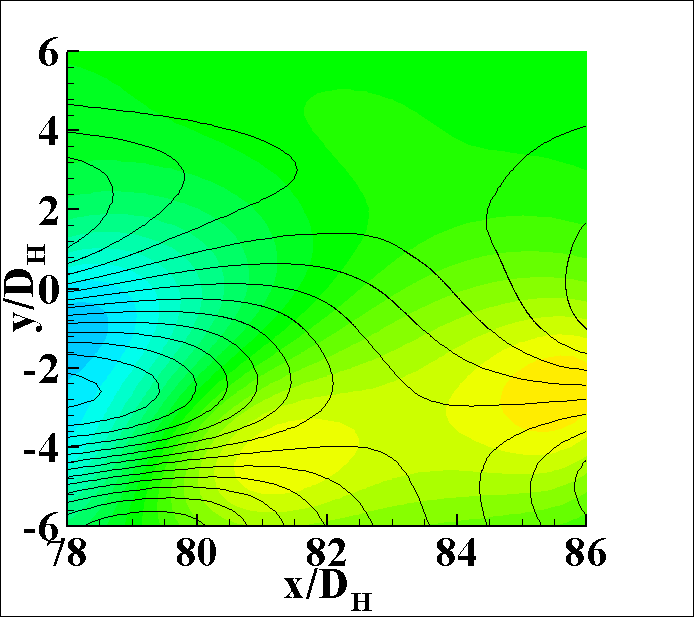}}\subfloat[\label{uvcont_31_460}] {\includegraphics[trim = 6 4 10 10,clip,width=0.19\textwidth]{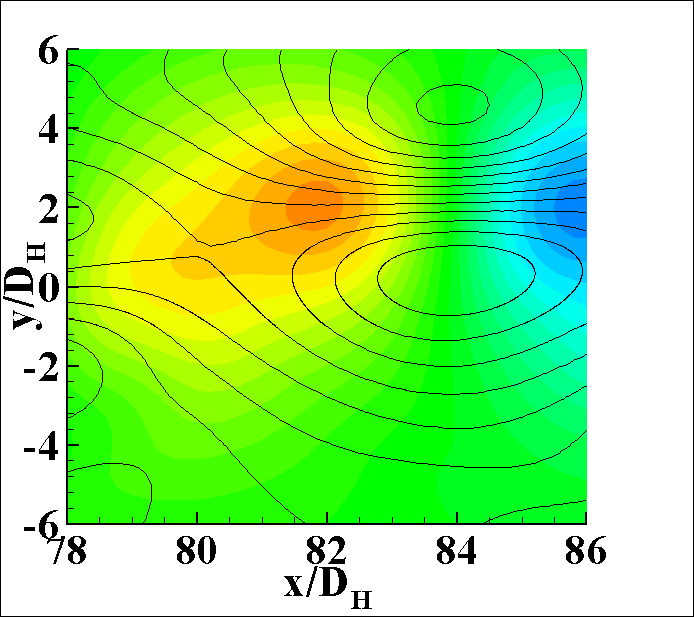}} \subfloat[\label{uvcont_31_740}] {\includegraphics[trim = 4 4 5 12,clip,width=0.22\textwidth]{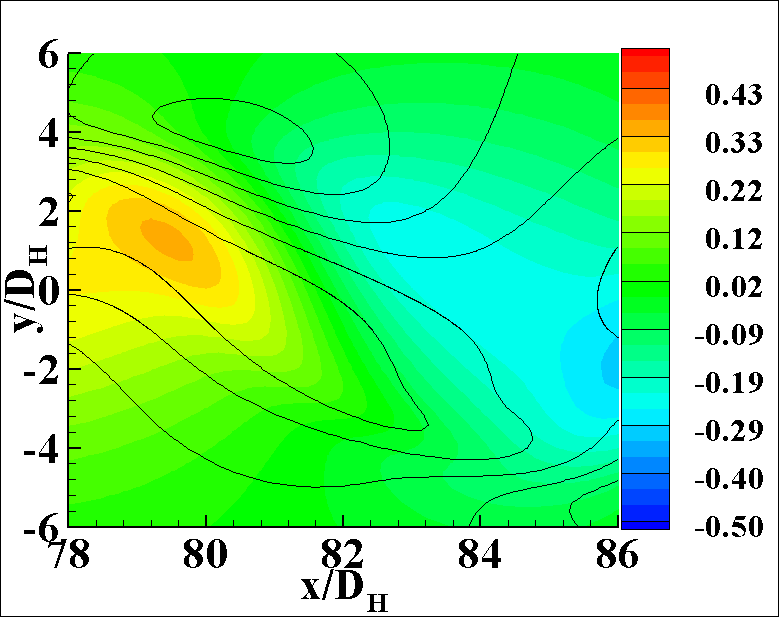}}}
\end{center}
\begin{center}
\mbox{\subfloat[\label{vort_tk}] {\includegraphics[trim = 8 4 10 10,clip,width=0.19\textwidth]{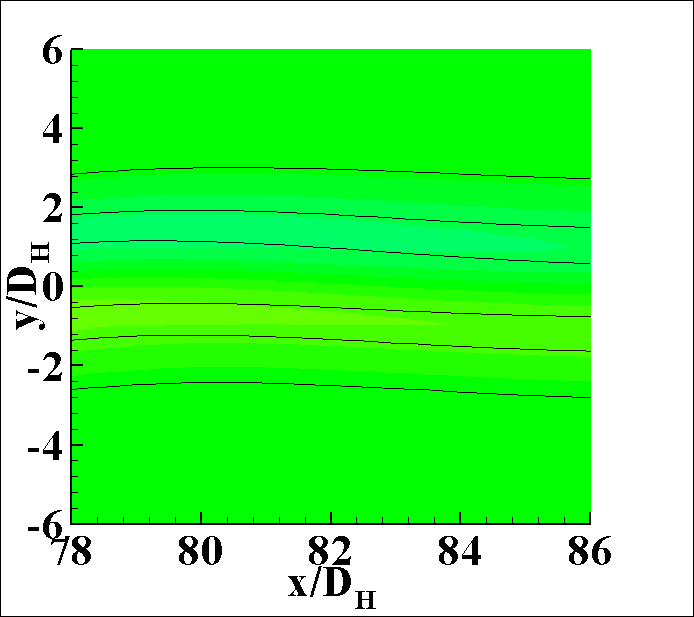}}\subfloat[\label{vort_tl}] {\includegraphics[trim = 8 4 10 10,clip,width=0.19\textwidth]{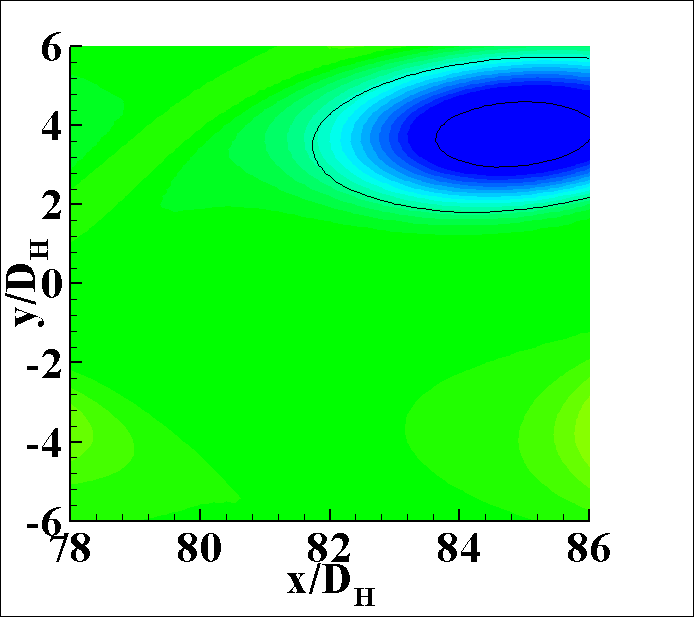}} \subfloat[\label{vort_tm}] {\includegraphics[trim = 8 4 10 10,clip,width=0.19\textwidth]{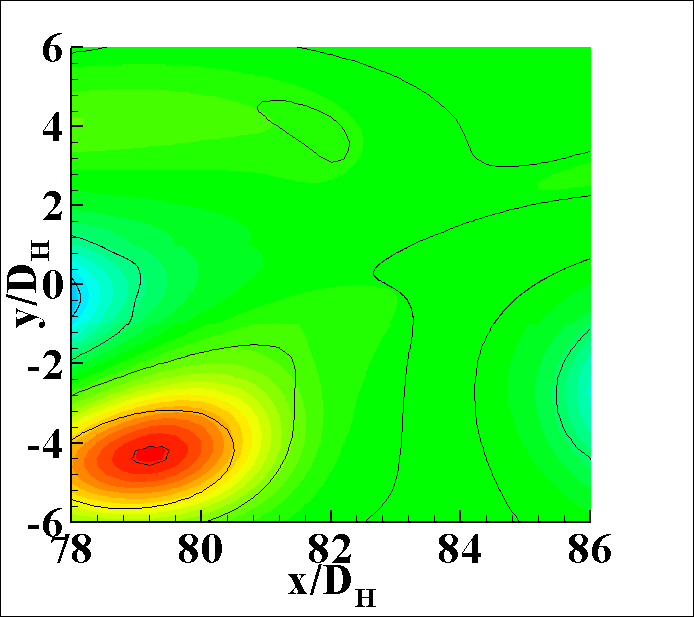}}\subfloat[\label{vort_tn}] {\includegraphics[trim = 8 4 10 10,clip,width=0.19\textwidth]{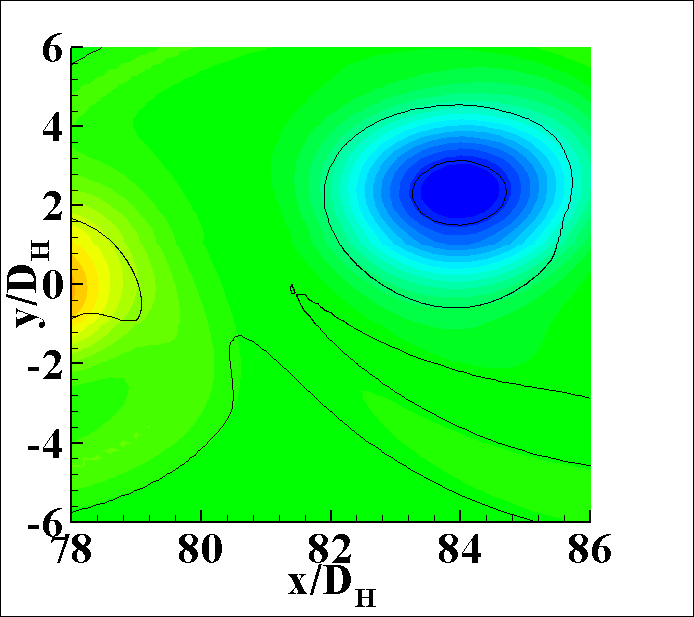}} \subfloat[\label{vort_to}] {\includegraphics[trim = 5 6 5 12,clip,width=0.22\textwidth]{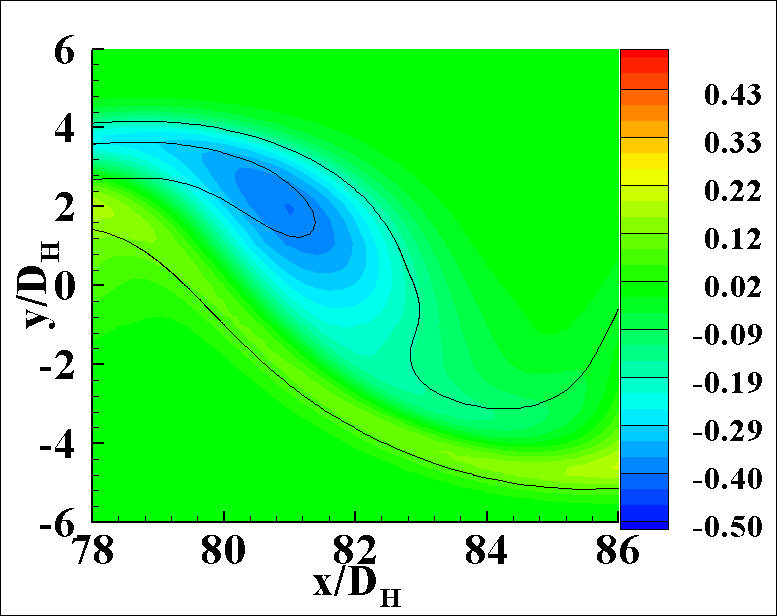}}}
\end{center}
\caption{ The temporal evolution of (a-e) u-contour lines(10 levels ranging from -0.4 to 1.6 (negative levels represented in dashed lines)) and v(flood) contours and (f-j) are temperature(20 levels ranging from 0.01 to 0.6) contour lines at ($x/D_{H}$,$y/D_{H}$) = (62,0) respectively}
\label{uvcont3}
\end{figure*}
Finally, in this subsection, we provide the physical sources of the frequencies observed in the spectra plotted for the secondary shedding region. Figure \ref{uvcont3} shows the temporal development of the wake around $x/D_H$=82 which belongs to the secondary shedding region. As observed in the signal decomposition method, the secondary shedding region does not have a saturated state of wake development. Therefore, all the sub-figures in Fig. \ref{uvcont3} represent the transition regime for the secondary shedding region.

Three important observations can be made in Fig. \ref{uvcont3}. Firstly, note the completely chaotic nature of the velocity and vortical structures. Here, the structures do not follow any periodic movements as we noted in the transition regime of the von-Karman and the parallel vortex regions. Secondly, the wavelength of these structures also seem to be larger than the structures of the von-Karman region. Finally, these structures also have diverse sizes. Due to these three reasons, the FFT of primitive variables of the secondary shedding region predominantly result in lower frequency with other higher frequencies less-defined. This explanation is applicable to all the primitive variables as chaotic nature is present in all of them.
\section{Summary and conclusions}\label{conc}
We studied the spectra of primitive variables of fluid flow and scalar field in the wake of an elliptic cylinder of axis ratio 0.4. The semi-major axis of the cylinder was kept perpendicular to the free-stream flow. The Reynolds number of the simulation is 130 for which the previous studies reported that the flow is purely two-dimensional. The Prandtl number of the simulation is 0.71. 

In order to establish the nature of the secondary vortex street, we ran the simulation for long enough to find out that the secondary shedding is a time dependent phenomenon. This means that the spatial location at which the secondary shedding occurs keeps moving towards the cylinder as the simulation time progresses. However, after a certain time, the secondary shedding becomes established.

We observed three distinct regions of wake  in the downstream of the cylinder. The von-Karman region exists in the immediate downstream of the cylinder where the vortices are arranged periodically. After this region, the vortices get stretched along the streamwise direction and they align parallel to each other. That is why we call this region the parallel vortex region. Finally, the parallel vortices merge again in the secondary shedding region.

We carried out fast Fourier transform of signals of primitive variables along the wake centerline. Interestingly, we observed the low-frequency unsteadiness even in the von-Karman region, which is believed to be insensitive to wake development for streamwise velocity, pressure and temperature. To our surprise, the cross-stream velocity did not exhibit any low-frequency unsteadiness in the von-Karman region. On the other hand, we noted a range of higher and lower frequencies in the spectra of all the primitive variables for the parallel vortex and secondary shedding regions of the wake. We also noted that the imprints of the temperature are reflected on the streamwise velocity and pressure, but not in the cross-stream velocity. Also, the primary frequency of the streamwise velocity is twice the cross-stream velocity.

We then went on to provide the spectral and physical origins of the low-frequency unsteadiness in primitive variables of fluid flow and scalar. In order to provide the spectral source of the low-frequency unsteadiness, we considered an improved method proposed by \citet{pauls16}. Based on this method, we first divided the temporal signal into different segments according to the complex demodulation technique. We found that the primary high frequency of the primitive variables in the von-Karman region is from both the transitional and temporal parts of the signal. On the other hand, in the parallel vortex region, the transitional part of the signal resulted in low-frequency while the saturated part of the signal yielded only the primary frequency. From this, we were able to accurately point out that the transitional regime of the signal is the spectral source of the low frequency unsteadiness. 

Finally, we used instantaneous contours of flow and scalar field in order to establish the physical source of low-frequency unsteadiness. We noted that the signed von-Karman flow structures are indeed sensitive to the wake development, in contrast to the observation noted from the signal decomposition method. Yet, the overall $v$-velocity structure does not change in the von-Karman region, and this is why no low-frequency unsteadiness is observed in the $v$-velocity for the von-Karman region. In the parallel vortex and secondary shedding regions, the temporal wake development aids chaotic movement of flow structures where their wavelength gets modified. Due to this, the transitional regime of the signal results in low-frequency unsteadiness. 

In summary, the spectral source of the low-frequency unsteadiness is the transitional part of the signal identified by increasing amplitude with respect to time. On the other hand, the chaotic behaviour of the flow structures that paves the way for the transmutation of their wavelengths is the physical source of the low-frequency unsteadiness.
\begin{acknowledgments}
The authors gratefully acknowledge the excellent computing facilities provided by High Performance Computing Center of IIT Madras. 
\end{acknowledgments}

\bibliography{mybibfile}
\end{document}